\documentclass[final,3p,times]{elsarticle}
\journal{Journal of Systems and Software}

\usepackage{hyperref}
\usepackage{microtype}

\newtheorem{definition}{Definition}
\usepackage{arydshln} 
\usepackage{caption}

\usepackage{booktabs}
\usepackage{multirow}
\usepackage{tabularx}
\usepackage{siunitx}  
\usepackage{threeparttable}  
  \usepackage[breakable, skins]{tcolorbox}

  \newtcolorbox{promptbox}[1][]{
    colback=white,
    colframe=black,
    fontupper=\ttfamily\footnotesize,
    boxrule=0.8pt,
    arc=0pt,
    left=2pt,
    right=2pt,
    top=2pt,
    bottom=2pt,
    breakable,
    enhanced,
    extras unbroken={},  
    extras first={frame code={\draw[black, line width=0.8pt]
      (frame.north west) -- (frame.south west)
      (frame.north east) -- (frame.south east)
      (frame.north west) -- (frame.north east);}},
    extras middle={frame code={\draw[black, line width=0.8pt]
      (frame.north west) -- (frame.south west)
      (frame.north east) -- (frame.south east);}},
    extras last={frame code={\draw[black, line width=0.8pt]
      (frame.north west) -- (frame.south west)
      (frame.north east) -- (frame.south east)
      (frame.south west) -- (frame.south east);}},
    #1
  }

\usepackage{tcolorbox}
\usepackage{xcolor}
\definecolor{skyback}{HTML}{D6EAF8}
\definecolor{darkblue}{HTML}{1A5276}

\newcommand{\takeaway}[2][Summary]{
  \begin{tcolorbox}[
    colback=skyback!70,
    colframe=skyback!70,
    rounded corners,
    boxrule=0pt,
    left=8pt, right=8pt, top=6pt, bottom=6pt
  ]
  \textbf{\color{darkblue}#1:} #2
  \end{tcolorbox}
}

\newcolumntype{Y}{>{\centering\arraybackslash}X}

\begin{document}

\begin{frontmatter}

\title{Adaptive Strategy Generation for Boundary Value Exploration Beyond Numeric Inputs} 

\author{Sabinakhon Akbarova\corref{cor1}\fnref{label1}}
\ead{sabina.akbarova@chalmers.se}
\author{Felix Dobslaw\fnref{label2}}
\ead{felix.dobslaw@miun.se}
\author{Robert Feldt\fnref{label1}}
\ead{robert.feldt@chalmers.se}
\cortext[cor1]{Corresponding Author}
\affiliation[label1]{organization={Department of Computer Science and Engineering, Chalmers University of Technology},
            city={Gothenburg},
            country={Sweden}}

\affiliation[label2]{organization={Mid Sweden University},
            city={Östersund},
            country={Sweden}}

\begin{abstract}

Software behavior often changes abruptly at boundaries between input regions, and these transitions are known to be fault-prone. Boundary Value Exploration (BVE) automates boundary discovery by searching for pairs of similar inputs that nevertheless trigger different program behaviors. Existing automated BVE techniques rely on mutation operators hand-engineered for each input type, or even for each function under test, which has confined their use to numeric inputs. We present ABEX, an agentic LLM-based framework that replaces operator engineering with adaptive strategy generation: specialized LLM agents propose, select, and execute boundary-exploration strategies, guided by execution feedback and a quality-diversity (QD) archive. Because strategies are expressed in natural language, they can encode both type-level and function-specific knowledge, and effective strategies can even be stored and reused. We evaluate ABEX in a black-box setting on 20 functions with numeric, string, array, and mixed inputs. On numeric functions, ABEX outperforms a state-of-the-art QD method on 10 of 11 functions, with average QD-scores 11.7x higher. On non-numeric functions, addressed here for the first time in automated black-box BVE, ABEX discovers domain-aligned boundary behaviors for all subjects. Mutation testing shows the discovered boundaries are fault-revealing: with equally sized test suites, ABEX reaches an average mutation score of 86.2\% versus 61.9\% for the QD baseline, and kills nine times as many hard-to-detect stubborn mutants. An ablation study identifies adaptive strategy generation as the primary driver of these gains.

\end{abstract}

\begin{keyword}
Boundary value exploration \sep automated software testing \sep large language models \sep quality-diversity optimization \sep AI4SE
\end{keyword}

\end{frontmatter}


\section{Introduction}

Software systems often change behavior at the borders between input regions. These transitions can be especially fault-prone, which makes boundary-focused testing an efficient way to find defects with relatively few test cases~\cite{white1980domain, clarke1982close, hierons2006avoiding}. Boundary Value Exploration (BVE) makes this idea concrete by casting boundary discovery as a search for pairs of nearby inputs that nevertheless trigger different program behaviors, thereby exposing behavioral boundaries~\cite{dobslaw2020boundary}.

Recent work has started to automate BVE. AutoBVA~\cite{dobslaw2023automated} searches for boundary-inducing input pairs by optimizing the program derivative (PD)~\cite{feldt2019towards}, a metric that rewards pairs of inputs that are close in the input space but produce different outputs. This works well for sharp transitions, but optimizing a single metric can pull the search into a narrow part of the space and miss boundary areas that score lower yet still matter for testing. SETBVE~\cite{akbarova2025setbve} addresses this limitation with a quality-diversity (QD) approach based on MAP-Elites~\cite{mouret2015illuminating}, maintaining an archive of boundary candidates that balances boundary strength against behavioral diversity.

Despite this progress, automated BVE approaches still face a fundamental limitation: they depend on mutation operators tailored to the input data types of the function under test (FUT) and, for effective exploration, often to the FUT’s specific semantics. Designing generic operators is straightforward for numeric inputs, where boundary candidates can be generated by incrementing, decrementing, or sampling around numeric thresholds. However, this becomes harder for other and more complex data types such as strings, arrays, or mixed inputs. For example, consider a function that validates email addresses. A generic string mutation operator may randomly insert, delete, or replace characters, but such mutations are unlikely to systematically explore the meaningful boundary cases of the email format. Inputs such as \texttt{alice@example.com}, \texttt{aliceexample.com}, \texttt{alice@@example.com}, and \texttt{alice@example..com} differ by small syntactic changes, yet they exercise distinct boundary conditions related to the presence and position of @, dots, local-part structure, and domain structure. Generating such cases requires knowledge of the expected input grammar and of the validation rules implemented by the FUT. Similar challenges arise for file paths, URLs, identifiers, JSON objects, or other structured inputs whose boundary behavior depends on domain-specific constraints. Each new type, argument combination, or function therefore requires bespoke operator engineering, which is labor-intensive, error-prone, and difficult to scale. As a result, automated BVE has remained largely confined to numeric domains with generic operators, even though most real software operates on richer input structures whose boundary behavior depends on function-specific logic. This limitation motivates our work, which aims to reduce the need for manually engineered, type- and function-specific mutation operators.

We present ABEX, an agentic LLM-based framework designed to remove this bottleneck. The central concept in ABEX is a boundary-exploration \textit{strategy}: a reusable, natural-language procedure that instructs the system how to generate candidate input pairs for a given FUT. A strategy is not a single test case or a fixed low-level mutation operator; rather, it specifies an exploration policy, such as which input structures to vary, whether to generate new pairs from scratch or mutate existing boundary candidates. For example, for an email-validation function, a strategy may instruct the system to vary the position and number of @ symbols, alter dots in the local part or domain, and pair syntactically similar valid and invalid addresses to expose format-validity boundaries. 

Building on work on autonomous testing agents~\cite{feldt2023towards, yoon2024intent} and Automated Design of Agentic Systems (ADAS)~\cite{hu2024adas}, ABEX combines a QD archive and the program derivative metric with specialized agents that propose, refine, select, and execute such strategies. Because these strategies are expressed in natural language, they can capture both type-level and FUT-specific knowledge derived from docstrings, specifications, or other semantic information without requiring changes to the framework. The resulting strategies are human-readable and, in principle, reusable: effective strategies discovered for one function or input type can be stored and applied to others, allowing the framework to accumulate testing knowledge over time. Although the use of LLMs leads to higher per-iteration cost than optimized numeric search, ABEX compensates by finding high-quality boundaries early and by working on input types that existing methods cannot handle.

We evaluate ABEX on 20 FUTs with integer, string, array, and mixed inputs. On numeric inputs, ABEX outperforms both the QD-based baseline SETBVE \cite{akbarova2025setbve} and a single-prompt LLM baseline on 10 of the 11 evaluated functions. Under the same budget as SETBVE, 100 iterations with 10 candidates generated per iteration, with a total of 1000 generated candidate input pairs, ABEX achieves an average QD-score $11.7\times$ higher. The single-prompt baseline serves as a check on whether simply asking an LLM for boundary candidates, without iteration, feedback, or agents, is already sufficient; following prior work \cite{guo2025boundary}, it generates 50 candidate pairs in a single call, and ABEX achieves an average QD-score $2.8\times$ higher (Section \ref{subsec:experimental_setup} details both budgets). For non-numeric inputs, which to the best of our knowledge are explored here for the first time in automated black-box BVE, ABEX discovers meaningful, domain-aligned boundary behaviors across all tested functions. Mutation testing further shows that the boundary candidates discovered by ABEX are effective as fault-detecting test cases: under equally sized test suites, ABEX achieves an average mutation score of 86.2\% on numeric functions, compared to 61.9\% for SETBVE, and kills roughly nine times as many stubborn mutants. For non-numeric functions, ABEX-generated suites reach an average mutation score of 76.3\%. An ablation study identifies adaptive strategy generation as the primary driver of these gains: on a six-FUT subset, the full framework achieves an average per-FUT QD-score improvement of $6.8\times$ over a variant that iterates a single fixed strategy.

The contributions of this paper are:
\begin{itemize}
    \item ABEX, an adaptive boundary exploration framework that autonomously generates and accumulates reusable exploration strategies, achieving $2.8\times$ and $11.7\times$ higher QD-scores than single-prompt LLM and QD-based baselines, respectively, on numeric inputs.
    \item The first automated BVE results in a black-box setting for non-numeric inputs: ABEX discovers domain-aligned boundary behaviors across all 10 string, array, and mixed-input functions.
    \item Mutation-testing evidence that the discovered boundary candidates are effective at detecting seeded faults: ABEX reaches an average mutation score of 86.2\% versus 61.9\% for the QD-based baseline on numeric FUTs, and kills roughly nine times as many hard-to-detect stubborn mutants.
    \item An open replication package\footnote{https://zenodo.org/records/21354853} containing the ABEX framework and evaluation dataset.
\end{itemize}

Section~\ref{sec:background} presents background and related work. Section~\ref{sec:framework} describes the ABEX architecture with a running example. Section~\ref{sec:evaluation} outlines the experimental setup. Section~\ref{sec:results} reports results, followed by discussion and threats to validity in Section \ref{sec:discussion}. Section~\ref{sec:conclusion} concludes.

\section{Background and Related Work} \label{sec:background}

This section introduces key concepts behind boundary discovery and reviews related work on quality-diversity optimization and LLM-based testing.

\subsection{Boundary Discovery}

Boundary Value Analysis (BVA) and Boundary Value Testing (BVT) focus on inputs near transitions between behavioral domains (equivalence partitions), where faults are more likely to occur~\cite{clarke1982close, hierons2006avoiding}. Traditionally, identifying such boundaries has been largely manual. Early work by White and Cohen proposed domain testing, which identifies boundaries between mutually exclusive subdomains to detect control-flow errors~\cite{white1980domain}. Subsequent efforts aimed to reduce the manual effort involved in boundary identification. For example, Jeng et al.~\cite{jeng1999automatic} introduced a semi-automated approach that combines dynamic search with algebraic manipulation of boundary conditions.

Fully automated approaches have also been proposed, but they typically rely on explicit program specifications or access to internal program structure. Pandita et al.\cite{pandita2010guided} presented a white-box technique that instruments code to identify boundary values near decision points, while Zhang et al.\cite{zhang2015boundary} leveraged symbolic execution to extract boundary conditions and generate test cases through constrained combinatorial testing. Although effective, these methods depend on well-defined partitions or specifications, which limits their applicability when such information is incomplete, ambiguous, or unavailable.

Boundary Value Exploration (BVE) was introduced to address this limitation by treating boundary discovery as a systematic search problem~\cite{dobslaw2020boundary}. AutoBVA~\cite{dobslaw2023automated}, the first automated black-box framework evaluated on numeric-input functions, uses the concept of the \emph{program derivative} (PD)~\cite{feldt2019towards} as a search objective. The PD measures how sensitive program behavior is to changes in input by computing the ratio of output distance to input distance\footnote{Input and output distances calculation depend on datatype (e.g., Euclidean for numeric values and Jaccard for strings).} $PD(i_1, i_2) = \frac{d_o(o_1, o_2)}{d_i(i_1, i_2)}$. Input pairs with $\text{PD} > 0$ are considered \emph{boundary candidates}, and larger PD values indicate sharper behavioral transitions.

While effective at identifying strong boundaries, optimizing solely for PD can bias the search toward high-contrast regions and overlook other relevant boundary areas with lower PD values due to program-specific characteristics~\cite{akbarova2025setbve}. Moreover, PD depends on the chosen distance metrics, and no single metric fully captures boundariness across domains. Consequently, PD values are not directly comparable across the entire input space, and smaller PD values in some regions may be more informative than larger ones in others.

Boundary testing beyond numeric inputs has received some attention. Jain et al. \cite{jain2010boundary} manually apply classical BVA schemes to character inputs by treating letters as ordered values within a known range, without automation or boundary discovery. Zhao et al. \cite{zhao2010automatic} automatically generate test points for string predicate borders, but their technique is white-box. Neither approach discovers behavioral boundaries of a black-box function whose partitions are unknown. Automated BVE for non-numeric inputs in a black-box setting has, to our knowledge, not been addressed.

\subsection{Quality-Diversity Optimization}

Quality-Diversity (QD) algorithms aim to discover diverse sets of high-performing solutions rather than a single optimum. A widely used QD algorithm is MAP-Elites, introduced by Mouret and Clune \cite{mouret2015illuminating}. MAP-Elites maintains a grid-based \emph{archive} of elite solutions organized by \emph{behavioral descriptors} that represent different regions of the behavioral space. Each cell stores the best solution discovered for that region.

QD methods have been applied to software testing across several domains. Marculescu et al.~\cite{marculescu2016using} showed that exploration-focused algorithms including MAP-Elites cover a broader portion of the behavior space than objective-based approaches, with over 80\% of their outputs not found by the objective-based method, demonstrating their suitability for high-dimensional behavioral spaces under limited resources. Feldt and Poulding~\cite{feldt2017searching} further investigated diversity-driven test input generation inspired by Novelty Search \cite{boussaa2015noveltysearch} and MAP-Elites. Xiang et al.~\cite{xiang2023automated} applied MAP-Elites to test suite generation for Software Product Lines. QD algorithms have also been applied to deep learning system testing: Riccio and Tonella~\cite{riccio2020model} developed DeepJanus to map the behavioral frontier of deep learning systems using input pairs with similar features but differing behaviors, and Zohdinasab et al.~\cite{zohdinasab2021deephyperion} developed DeepHyperion, which uses Illumination Search to explore interpretable feature spaces of deep learning inputs. More recently, Weißl et al. \cite{weissl2026targeted} proposed Mimicry, which shows that targeted boundary exploration outperforms the untargeted perturbations used by earlier deep learning testing techniques. While some of these tools operate in a black-box manner, they still assume specific input/output types and domain-specific quality signals~\cite{riccio2020model}, limiting their generalization to traditional software with heterogeneous input types.

Most closely related to our work, Akbarova et al. proposed SETBVE~\cite{akbarova2025setbve}, a QD-based framework for automated BVE. Their results show that balancing solution quality with behavioral diversity enables the discovery of a wider range of edge-case behaviors than approaches that optimize for quality alone. SETBVE guides the search using a MAP-Elites archive defined by four behavioral descriptors derived from the input and output characteristics of a FUT. ABEX builds on this formulation by adopting the same behavioral descriptors.

\subsection{Software Testing with LLMs}

Large Language Models (LLMs) have recently demonstrated strong capabilities in software engineering tasks, including automated testing \cite{nass2024improving, lemieux2023codamosa, feng2024prompting}. A recent survey by Wang et al.~\cite{wang2024software} reviews over 100 studies applying LLMs to software testing, highlighting their effectiveness across tasks such as test case generation, debugging, and program repair. These works show that LLMs are particularly well-suited for generating test inputs by leveraging their understanding of programming patterns and semantics \cite{boukhlif2024towards, chen2024chatunitest}.

However, most existing approaches rely on single-shot generation or fixed prompting schemes without incorporating execution feedback \cite{guo2025boundary, pan2025aster}, which might lead to redundant, low-quality, or invalid test cases \cite{gu2024testart}. While some recent work explores iterative or feedback-driven testing \cite{altmayer2025coverup}, these approaches typically operate within predefined generation patterns and do not adapt how the input space is explored over time.

Agent-based systems have recently emerged as a promising paradigm for structuring LLM-driven workflows. In software testing, DroidAgent introduced a multi-agent architecture for Android GUI testing, where agents maintain system knowledge and generate high-level exploration goals \cite{yoon2024intent}, building on earlier work on LLM-based autonomous testing agents \cite{feldt2023towards}.

Inspired by these advances, ABEX integrates LLM-driven strategy generation within a quality-diversity search framework, enabling adaptive boundary discovery guided by dynamically generated exploration strategies.

\section{Framework} \label{sec:framework}

\begin{figure}[]
    \centering
    \includegraphics[width=0.6\linewidth]{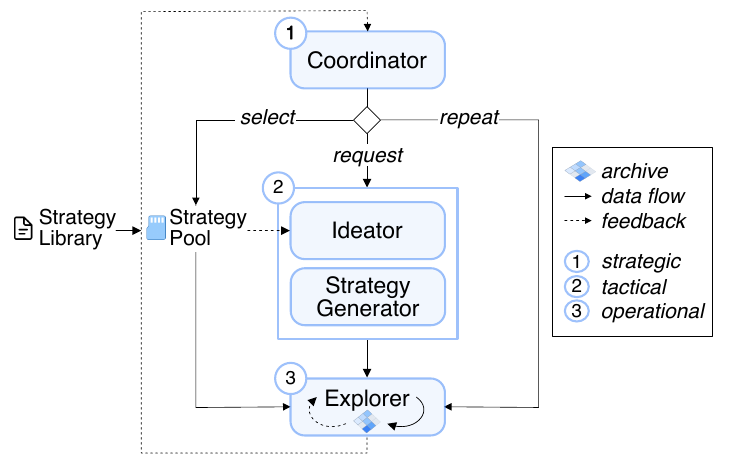}
    \caption{High-level overview of ABEX architecture. The Coordinator (strategic layer) decides whether to repeat, select, or request a strategy; the Ideator and Strategy Generator (tactical layer) create new strategies; the Explorer (operational layer) executes strategies and updates the MAP-Elites archive.}
    \label{fig:overview_abex}
\end{figure}

ABEX follows a three-layer agent architecture inspired by hierarchical frameworks in management science that separate strategic, managerial (tactical), and operational levels~\cite{anthony1965planning} (see Figure~\ref{fig:overview_abex}). Strategic decisions in (1) determine how the input space is explored. Tactical decisions in (2) materialize as the construction or selection of an input space exploration strategy. Finally, the operational layer (3) generates candidate inputs according to the chosen strategy and executes them on the FUT. This separation supports efficient resource allocation and allows different LLM configurations to be assigned to each layer~\cite{wu2023autogen}: strategic decisions are infrequent but high-impact, while operational actions are frequent and constrained. ABEX iterates in a loop until the search budget is exhausted.

The following subsections describe the role of each component and walk through one iteration using a running example on a \texttt{date} FUT, taking three integer inputs (\texttt{year}, \texttt{month}, \texttt{day}) and returning an ISO date string \footnote{Prompt examples are truncated for readability and complete prompts are available in the replication package.}.

\subsection{Coordinator} \label{sec:coordinator}

The \emph{Coordinator} implements the strategic layer of ABEX and directs the overall exploration process. Exploration is driven by \emph{strategies} --- textual, step-by-step instructions for generating candidate input pairs. During the search, available strategies are maintained in a \emph{strategy pool}, which is initialised from a \emph{strategy library}: an offline repository of reusable strategies. In this study, the library contains a single pre-seeded \emph{baseline strategy} that instructs the system to ``generate diverse boundary value input pairs for the given function''. This baseline receives the FUT signature and docstring --- a one- to two-sentence natural-language description of the function's purpose --- but specifies no structured procedure. Since it receives the signature and docstring, it is formally a \emph{context-aware generation} strategy (see Section \ref{subsec:strategy_generator} for more details), but we track it separately in our analysis (Section \ref{sec:rq3}) because it is pre-seeded rather than produced by the tactical layer during the search. The first iteration executes this baseline strategy directly. The Coordinator is invoked only after the active strategy's stopping condition is met (Section~\ref{subsec:explorer}).

Once activated, the Coordinator decides among three exploration directions. First, it may continue using the most recently executed strategy if it appears promising or insufficiently explored. Second, it may switch to another strategy from the strategy pool, selecting the specific strategy to activate. Third, it may request the creation of a new strategy by specifying the desired strategy \textit{mode}, \textit{type}, and the \textit{search gap} that the strategy should address (Section~\ref{subsec:strategy_generator}).

To support this decision-making process, the Coordinator receives contextual information about the search. This includes the current iteration number, allowing it to reason about the stage of exploration, as well as information about the FUT, such as its signature and docstring. The Coordinator also has access to tools that summarize search progress, such as recent archive updates and stagnation indicators. Stagnation occurs when no new archive cells are discovered and no improvements are observed in existing cells for several consecutive iterations. In addition, the Coordinator can inspect the pool of available strategies, including a short description of each strategy and its usage statistics. These statistics include both total success rate (new boundary candidates and quality improvements over all attempts) and a time-decayed success rate that weights recent performance more heavily. Comparing these two metrics enables the Coordinator to assess not only a strategy's overall effectiveness but also whether its performance is improving or declining.

This adaptive decision process allows ABEX to dynamically adjust its search behavior as exploration progresses, either by revisiting strategies that may still be promising or by requesting new strategies when experiencing stagnation.

\begin{promptbox}
\footnotesize
\textbf{Coordinator} \\[4pt]
  \texttt{[System]} Choose the best next action to maximize the discovery of diverse boundary test cases: request new strategy, repeat current strategy or select another available strategy. [...truncated...]\\

  \texttt{[Input]}\\
  Iteration progress: Iteration 12 of 100\\
  Function signature: date(year: int, month: int, day: int) -> str\\

  \texttt{[Coordinator calls tool: stagnation\_check]}\\
  Progress slowing - 3 consecutive rounds with no progress...\\

  \texttt{[Coordinator calls tool: strategies]}\\
  Strategies sorted by recent success rate:\\

  -{-}- ID: strategy1 (CURRENT) -{-}-\\
  Summary: Generate pairs testing integer boundary values like 0, -1, MAX\_INT.\\
  Mode: generation, Type: generic\\
  Used in 8 of 11 past rounds. Last used 1 rounds ago.\\
  Success rate: 0.31, Recent success rate: 0.23\\

  -{-}- ID: strategy2 -{-}-\\
  Summary: Test month-day combinations at month boundaries (30/31 days, February).\\
  Mode: generation, Type: context-aware\\
  Used in 3 of 11 past rounds. Last used 4 rounds ago.\\
  Success rate: 0.22, Recent success rate: 0.2\\

  \texttt{[Output]}\\
  option: request\\strategy\_mode: generation\\strategy\_type: context-aware\\search\_gap: target year boundary transitions affecting leap years.
  \end{promptbox}

\subsection{Ideator and Strategy Generator}
\label{subsec:strategy_generator}

When a new strategy is required, the tactical layer generates one through the \emph{Ideator} and the \emph{Strategy Generator}. The Coordinator specifies three elements for the new strategy: the strategy \textit{mode}, its \textit{type}, and the \textit{search gap} that the strategy should target.

\begin{definition}[Strategy] \label{def:strategy}

\label{def:strategy}

A strategy \(s\) is a tuple $s = (m,t,P)$, where \(m \in \{\mathrm{generation}, \mathrm{mutation}\}\) is the strategy mode,
\(t \in \{\mathrm{generic}, \mathrm{context\mbox{-}aware}\}\) is the strategy type, and $P = \langle p_1, p_2, \ldots, p_k \rangle$ is an ordered sequence of natural-language steps that, given a function under test \(f\) and optionally a sampled input pair $a = (i_1,i_2)$ from the archive $A$, produces a set of candidate input pairs $\{(i_1^j,i_2^j)\}_{j=1}^{N}.$ A generation strategy requires no sampled input pair, i.e., \(a=\mathrm{none}\), while a mutation strategy requires \(a \in A\).

\end{definition}

The mode \(m\) determines how input pairs are obtained. Generation strategies produce new input pairs from scratch using the available FUT information. Mutation strategies instead operate on existing archive entries: they first sample\footnote{An archive cell is sampled uniformly at random.} an input pair from the archive (each archive cell stores one pair), and then apply small modifications to explore nearby regions of the input space.

The type \(t\) determines what information the generation strategy may use; mutation strategies never receive the function signature or docstring and are therefore always of type generic. \textit{Generic} generation strategies only know the number and data types of the input arguments, and therefore rely on general boundary heuristics. In contrast, \textit{context-aware} generation strategies additionally have access to the function signature and its docstring. This enables the strategy to exploit domain-specific hints about the input space.

For example, if the LLM only knows that a function accepts three integers, it may explore general numeric boundaries such as \texttt{INT\_MAX} or \texttt{INT\_MIN}. However, if it also knows that the function is \texttt{date(year, month, day)}, it may instead focus on domain-specific transitions such as month changes or leap years. Both perspectives are useful for discovering different types of boundary behaviors.

The Ideator bridges the Coordinator's search gap and the executable strategy by
producing a tactical \textit{idea}: a one- to two-sentence conceptual direction that is
more specific than a search gap but not yet a full strategy. Separating idea generation from strategy construction serves two purposes. First, it identifies a promising strategy direction, or, when a
suitable direction already exists, selects an existing strategy rather than
producing a redundant one. Second, it filters candidate directions that are
likely to be low-yield before they are expanded into full strategies.

The first purpose requires comparing candidate directions by the boundary behavior they are expected to explore, not only by the wording of their descriptions. Lexical or embedding-based similarity checks can identify strategies that are close in phrasing, but they may miss functional redundancy between differently worded directions. For example,
\texttt{"integer overflow boundaries"} and \texttt{"values near MAX\_INT"} may appear different as text, yet both target essentially the same numeric boundary. At the same time, textually similar directions can still be useful if they target
different boundary conditions: \texttt{"test empty string input"} and \texttt{"test single-character input"} are close in wording but exercise different string-length boundaries. The Ideator therefore evaluates whether a candidate direction addresses the Coordinator's requested search gap and adds functional novelty relative to the existing strategy pool, rather than merely checking whether its wording differs.

The second purpose concerns the expected usefulness of a candidate direction before it is expanded into a full strategy. A direction may be novel but still low-yield if it is too broad to guide concrete input-pair generation, too narrow to produce diverse candidates, inconsistent with the requested generation or mutation mode, or likely to spend the search budget on invalid or behaviorally uninformative inputs. To filter such cases, the Ideator is instructed, within a single LLM call, to consider two to three candidate directions, compare them against the existing strategy pool in terms of search-gap relevance, functional novelty, and expected yield, and return the most promising one. 


Once an idea is selected, the Strategy Generator expands it into an executable exploration procedure by converting it into concrete steps describing how input pairs should be generated. The Strategy Generator also specifies the required output format and introduces variation within the steps so that repeated executions of the strategy produce diverse boundary candidates. The resulting strategy is then passed to the operational layer for execution.

  \begin{promptbox}
  \footnotesize
  \textbf{Ideator} \\[4pt]
  \texttt{[System]} Return one worthwhile new idea that is different from existing strategies and likely to discover new diverse boundary test cases. If none is worthwhile, return a fallback strategy. [...truncated...]\\

  \texttt{[Input]}\\
  existing\_strategies:\\
  -{-}- ID: strategy1 -{-}-\\
  Mode: generation, Type: generic\\
  Steps:\\
  1. Pick a boundary value (0, 1, -1, MAX\_INT, MIN\_INT). \\2. Create i1 at the boundary. \\3. Create i2 by adding or subtracting 1. \\...
  
  -{-}- ID: strategy2 -{-}-\\
  Mode: generation, Type: context-aware\\
  Steps: \\1. Select a month with 30 or 31 days. \\2. Set day to the last valid day. \\3. Create i2 by incrementing day by 1. \\...\\

  function\_signature: date(year: int, month: int, day: int) -> str\\
  requested\_strategy\_mode: generation\\
  requested\_strategy\_type: context-aware\\
  search\_gap: target year boundary transitions affecting leap years.\\

  \texttt{[Output]}\\
  Idea: \\
  Focus on year values where leap year rules change: years divisible by 4, 100, and 400, testing February 28/29 transitions across these boundaries.
  \end{promptbox}

  \begin{promptbox}
  \footnotesize
  \textbf{Strategy Generator} \\[4pt]
  \texttt{[System]} Turn the given idea description into a concise strategy for generating boundary test cases. [...truncated...]\\

  \texttt{[Input]}\\
  requested\_strategy\_mode: generation\\
  requested\_strategy\_type: context-aware\\
  function\_signature: date(year: int, month: int, day: int) -> str\\
  search\_gap: Target year boundary transitions affecting leap years.\\
  idea: Focus on year values where leap year rules change: years divisible by 4, 100, and 400, testing February 28/29 transitions across these boundaries.\\

  \texttt{[Output]}\\
  Strategy:\\
  1. Pick a leap year boundary: (a) divisible by 4, (b) divisible by 100, (c) divisible by 400.\\
  2. Set month = 2 and day = 28 or 29.\\
  3. For i1, use a year just before the boundary; for i2, use a year at or after.\\
  4. Vary across boundary types and day values.\\
  5. Output as \{"i1": [year, month, day], "i2": [year, month, day]\}
  \end{promptbox}
\subsection{Explorer} \label{subsec:explorer}

The \emph{Explorer} implements the operational layer of ABEX and executes the strategies produced by the tactical layer; the name follows the similar component in SETBVE with a comparable role~\cite{akbarova2025setbve}. Given a textual strategy, the Explorer invokes the LLM to generate $N$ input pairs per call\footnote{In this study, $N=10$.}. Each pair represents a potential boundary candidate and is evaluated using a MAP-Elites style archive.

The archive follows a QD formulation similar to SETBVE. It is organized as a four-dimensional grid in which each occupied cell corresponds to an exact integer coordinate in the \textit{behavioral descriptor} space. These descriptors capture observable program behaviors induced by an input pair. In this work, we adopt four descriptors~\cite{akbarova2025setbve}: the number of exceptions triggered by the pair, the output length difference, the total length of the stringified input pair, and the variance of the input lengths. We use the same four descriptors for all FUTs because they are type-agnostic: they depend on exceptions and stringified input/output properties rather than datatype-specific structure. Using this common descriptor space for both numeric and non-numeric FUTs also keeps archive-level measurements comparable across FUTs.

As in SETBVE, ABEX applies no additional binning to these descriptors: each distinct integer descriptor vector defines a separate archive cell. For example, two candidates mapped to \((0,3,12,2)\) and \((0,4,12,2)\) occupy different cells rather than being coarsened into the same interval. We avoid such coarsening because small changes in output length, input
length, or exception behavior can correspond to meaningful boundary differences. The archive is therefore implemented as a dynamically growing map: cells are created only when new descriptor coordinates are observed.

Descriptor values are integer-valued, and their observed ranges are FUT-dependent: the exception-count descriptor takes values in \(\{0,1,2\}\), while output-length difference, total input length, and rounded input-length variance are bounded below by \(0\) and, in principle, unbounded above. In practice, however, their observed ranges are limited by the generated inputs and the outputs produced by the FUT during a run.

Each generated or mutated input pair is assigned to a corresponding archive cell based on the behavioral descriptors by executing the FUT on each input in the pair. Consider the pair $i_1=\texttt{"a@b.co"}$ and $i_2 = \texttt{"a@b.c"}$ for the \texttt{validateEmail} FUT, which yields the outputs \texttt{"Valid"} and \texttt{"ValueError: Invalid email format."} respectively. This is a boundary where removing one character from the domain extension (\texttt{.co} $\to$ \texttt{.c}) makes the email invalid. The behavioral descriptors are computed as follows: (1) one output raises an exception giving an exception-count descriptor of 1; (2) the output lengths differ by $|5 - 33| = 28$; (3) the input lengths sum to $6 + 5 = 11$; and (4) the variance of input lengths is $\lfloor 0.5 \rfloor = 0$. The pair is therefore assigned to the cell at coordinates (1, 28, 11, 0). A different boundary such as \texttt{"user@example.com"} and \texttt{"userAexample.com"}, where replacing the \texttt{@} symbol with a letter removes the required delimiter, maps to a different cell at (1, 28, 32, 0) due to the longer total input length ($16 + 16 = 32$).

The quality of the input pair is then evaluated using the program derivative (PD), computed following SETBVE. Output distances are computed using Jaccard distance on stringified outputs, which both aligns with SETBVE and makes the metric applicable across all output types in our study. For inputs, we use Euclidean distance when all arguments are numeric, and an average of per-element distances otherwise,  using Jaccard distance for strings and arrays (stringified), and equality for numeric values (0 if identical, 1 otherwise)\footnote{We use bounded $[0,1]$ equality for numeric elements in mixed inputs to match Jaccard's range; unbounded Euclidean would otherwise dominate the average.}. The archive is updated following the MAP-Elites policy: if the cell is empty the input pair is inserted, if the cell is occupied but the new pair has higher quality (PD) it replaces the input pair occupying the cell, otherwise the archive remains unchanged.

The Explorer repeatedly executes the same strategy until a stopping condition is reached, either after 10 iterations or after three consecutive iterations without archive updates, meaning that no new cells are discovered and no existing cells are improved. The iteration limit encourages exploration diversity under a limited search budget by allowing other strategies to be attempted. If a strategy remains effective, the Coordinator may later select it again. To further encourage diversity, the Explorer is instructed to avoid generating input pairs that already exist in the archive. The prompt therefore includes a sample of previously discovered pairs together with their input-related properties, consisting of both the most recently added pairs and randomly selected ones from the archive\footnote{We include 10 recent and 30 randomly sampled input pairs; including the full archive would make the prompt excessively long.}. Once a stopping condition is met, control returns to the Coordinator together with the updated archive statistics, which are used to determine the next exploration step.

\begin{promptbox}
\footnotesize
\textbf{Explorer} \\[4pt]
\texttt{[System]} Generate 10 diverse input pairs (i1 and i2) by following the given strategy. [...truncated...]\\

\texttt{[Input]}\\
function\_signature: date(year: int, month: int, day: int) -> str\\

Strategy:\\
1. Pick a leap year boundary: (a) divisible by 4, (b) divisible by 100, (c) divisible by 400.\\
2. Set month = 2 and day = 28 or 29.\\
3. For i1, use a year just before the boundary; for i2, use a year at or after.\\
4. Vary across boundary types and day values.\\
5. Output as \{"i1": [year, month, day], "i2": [year, month, day]\}\\

Avoid existing input pairs:\\
i1: [2023, 1, 31], i2: [2023, 2, 1] (small length, similar magnitudes)\\
i1: [2023, 12, 31], i2: [2024, 1, 1] (small length, similar magnitudes)...\\

\texttt{[Output]}\\
\{"i1": [1999, 2, 28], "i2": [2000, 2, 28]\}\\
\{"i1": [1900, 2, 28], "i2": [1900, 2, 29]\} ...
\end{promptbox}

\section{Evaluation} \label{sec:evaluation}

This section presents the experimental setup used to evaluate ABEX.

\subsection{Research Questions}

Our evaluation addresses the following research questions.

\textit{RQ1: How effective are LLM-generated strategies at discovering boundaries for functions with numeric inputs?} This question evaluates whether LLM-generated strategies can discover boundary behaviors for FUTs with integer inputs. Integer FUTs enable direct comparison with the existing QD-based automated BVE framework. Effectiveness is measured using both the quality of discovered boundary candidates and the diversity of behaviors represented in the archive.

\textit{RQ2: Can ABEX discover boundaries for functions with non-numeric inputs, and what types of boundaries are identified?} This question investigates whether ABEX generalizes beyond numeric domains. We evaluate FUTs with string, array and mixed-type inputs and analyze the types of behavioral transitions discovered (e.g., domain-specific output properties).

\textit{RQ3: How do different exploration strategies influence boundary discovery across input data types?} This question analyzes the relationship between strategy characteristics and exploration performance. Specifically, we examine the usage frequency and effectiveness of different strategy modes and types across three FUT groups: numeric, string, and array or mixed inputs.

\textit{RQ4: How effective are boundary candidates discovered by ABEX at detecting faults compared to SETBVE?} This question evaluates fault-detection effectiveness using mutation testing. For FUTs with numeric inputs, we compare mutation scores and the number of uniquely killed mutants between test suites generated from ABEX- and SETBVE-discovered boundary candidates. For non-numeric FUTs, where SETBVE is not applicable, we report mutation testing results for ABEX without a baseline.

\textit{RQ5: How do ABEX's components and underlying LLM affect its ability to discover boundary candidates?}
This question evaluates the contribution of individual ABEX components through an ablation study and examines how the choice of closed (proprietary) or open-weight LLM affects boundary candidate generation capability.

\subsection{Experimental Setup} \label{subsec:experimental_setup}

\paragraph{Subjects}

Table~\ref{tab:futs} lists the 20 FUTs used in our evaluation. The set includes 10 functions with integer inputs, 5 with string inputs, 3 with array inputs, and 2 with mixed input types (array--integer and integer--boolean). The subject set was chosen to provide a controlled benchmark that still exercises different boundary-search scenarios. In particular, the FUTs vary not only in input datatype and arity, but also in the form of the observable behavior that defines boundaries: the benchmark includes categorical classifiers, format checkers, string transformation functions, array transformation functions, and search-like functions with input preconditions. 

One function, \texttt{complexCheck}, is evaluated in two variants: one with a one- or two-sentence docstring consistent with the other FUTs and one with a more detailed description (\texttt{complexCheck\_full}). This function is commonly used in white-box testing due to its multiple branching conditions with arbitrary thresholds; the richer description allows us to assess whether additional semantic information improves boundary discovery in our black-box setting.

We prioritize functions from prior BVE and testing literature where available. Since automated BVE in a black-box setting has not been explored for string inputs and only minimally for arrays, we include additional FUTs to cover a broader range of program logic. All FUTs were implemented in Python, closely following the original implementations where available.

\begin{table}
\centering
\renewcommand{\arraystretch}{1.2}
\footnotesize
\caption{Subject programs used in evaluation.}
\label{tab:futs}
\setlength{\tabcolsep}{2pt}
\begin{tabularx}{\columnwidth}{X X c c}
\toprule
\textbf{Name} & \textbf{Description} & \textbf{Number of arguments} & \textbf{Input type}\\
\midrule
bmi \cite{dobslaw2023automated, akbarova2025setbve} & Classify BMI category & 2 & Integer \\
bytecount \cite{dobslaw2023automated, akbarova2025setbve} & Format bytes with SI prefix & 1 & Integer\\
calDate \cite{zhang2015boundary} & Convert to Julian day number & 3 & Integer \\
circle \cite{akbarova2025setbve} & Point inside/outside circle & 2 & Integer \\
complexCheck$^{\dagger}$ \cite{zhang2015boundary, ghani2009searching} & Nested conditional branches & 3 & Integer \\
date \cite{dobslaw2023automated, akbarova2025setbve} & Format as ISO date string & 3 & Integer \\
english \cite{guo2024optimal} & Determine pass/fail status & 2 & Integer \\
findMiddle \cite{zhang2015boundary} & Return median of three & 3 & Integer \\
nextDate \cite{zhang2015boundary} & Compute next calendar date & 3 & Integer \\
tritype \cite{williams2005pathcrawler, zhang2015boundary} & Classify triangle type & 3 & Integer \\
passwordStrength & Assign strength level & 1 & String \\
printTokens & Tokens classification & 1 & String \\
replace & Match pattern and substitute & 3 & String \\
stringPalindrome \cite{jain2010boundary} & Check palindrome or not & 1 & String \\
validateEmail \cite{akbarova2026understanding} & Email format validity & 1 & String\\
insertionSort & Insertion sort algorithm & 1 & Array[int] \\
normalize \cite{akbarova2025setbve} & Min-max array scale to [0, 1] & 1 & Array[int] \\
maxLexString \cite{zhao2010automatic} & Max lexicographical string & 1 & Array[str] \\
binarySearch & Search for target value & 2 & Mixed (arr[int], int) \\
tcas \cite{zhang2015boundary} & Collision avoidance system & 12 & Mixed (int, bool) \\
\bottomrule
\end{tabularx}
\vspace{0.3em}
\footnotesize{$^{\dagger}$ The FUT has two specification versions (docstrings): a short version consistent with other FUTs and a more detailed one.}
\end{table}

\paragraph{Metrics}
Since ABEX maintains a MAP-Elites archive, our primary metric follows the QD paradigm. We report the \emph{QD-score}~\cite{pugh2016quality}, which measures both the quality and diversity of the archive population. Diversity is represented by the number of occupied archive cells $c$, while quality is given by the fitness values stored in these cells. The QD-score is defined as $\sum_{i=1}^{c} Q_i$, where $Q_i$ is the fitness value stored in cell $i$. In our setting, fitness corresponds to the program derivative (PD) of the input pair stored in the cell.

Because the QD-score does not explicitly indicate the number of discovered boundary candidates, we additionally report the number of archive cells that contain boundary candidates (i.e., archive cells with $\text{PD} > 0$), referred to as the \emph{BC count}.

\paragraph{Baselines}
For RQ1, we compare ABEX with two baselines: SETBVE and a single-prompt LLM. All experiments for RQ1 and RQ2 are repeated 10 times per configuration to account for stochastic variation.

SETBVE's original implementation evaluates runs using fixed wall-clock budgets (30 seconds or 10 minutes). Because LLM-based exploration incurs higher computation time per iteration, we instead compare methods using a fixed iteration budget\footnote{We use the open-source implementation and modify it to support an iteration-based budget.}. Specifically, SETBVE is executed for 1000 iterations, producing 1000 candidate input pairs. The corresponding ABEX configuration is run for 100 iterations, with each Explorer iteration generating 10 input pairs, also yielding 1000 candidate input pairs. We use the default SETBVE configuration, where initial sampling occupies 10\% of the total budget and mutation selects archive entries uniformly at random. ABEX follows the same uniform mutation policy.

The single-prompt baseline is a check on whether simply asking an LLM to generate boundary candidates, without iteration, feedback, or agents, is already sufficient. It uses the same LLM model and baseline strategy as the Explorer in ABEX, but without archive feedback or adaptive strategy selection. The baseline receives the function signature and docstring, and the model is prompted once to generate a diversified set of 50 boundary input pairs $(i_1, i_2)$, following the single-prompt setting of prior work~\cite{guo2025boundary}. The ExplorerOnly ablation (Section \ref{subsec:ablation}) complements this baseline: it iterates the same fixed strategy with archive feedback under a full search budget, which allows us to separate the effect of budget from the effect of adaptive strategy generation.

\paragraph{Large Language Models}
We employ a heterogeneous model configuration across components \cite{ye2025x}, assigning models based on cognitive demand and call frequency. The frequently invoked Explorer uses \texttt{gpt-5-mini} for cost-efficient instruction following, while the Coordinator and Ideator use reasoning-enhanced \texttt{gpt-5.1} for strategic decisions and idea generation. The Strategy Generator uses \texttt{gpt-5-mini} to convert ideas into executable strategies. Sampling temperatures are also role-dependent: the Coordinator and Strategy Generator use a moderate temperature (0.5) for moderately stable decisions, whereas the Ideator and Explorer use higher temperatures (0.8) to promote diversity in ideas and generated input pairs. This temperature selection is consistent with the findings of Lemieux et al.~\cite{lemieux2023codamosa}, which show that higher sampling temperatures have a consistently positive effect on exploration effectiveness. Section \ref{subsec:ablation} also evaluates two ABEX configurations based on the open-weight models Gemma-4 and Qwen-3.5.

ABEX is implemented in Python using the DSPy framework~\cite{khattab2023dspy}, which enables modular LLM pipelines through declarative task specifications (signatures). DSPy replaces manually engineered prompts with a prompt-programming approach that automatically compiles and optimizes prompts for each pipeline component, simplifying the development of multi-step LLM systems. All experiments were conducted on a MacBook Pro equipped with an Apple M2 Pro chip, 16GB RAM, running macOS 26.3.

\paragraph{Mutation Analysis}
To answer RQ4, we evaluate whether the boundary candidates discovered by ABEX and SETBVE are effective as test cases for detecting faults. We do this using mutation testing. Mutation testing creates faulty versions of the program under test, called mutants, by applying small syntactic changes to the original program. A test case kills a mutant if it produces a different output on the mutant than on the original program. Otherwise, the mutant remains live. We summarize test-suite effectiveness using the mutation score (MS), defined as the ratio of killed mutants to the total number of mutants.

For each FUT, mutants were generated once using \texttt{mutmut}~\cite{mutmut}, a Python mutation testing tool that applies mutation operators such as arithmetic, relational, and logical operator replacements. The same mutant set was then reused for all runs and compared frameworks. This ensures that the mutation results are based on the same faults, so differences in mutation score reflect differences in the generated test suites rather than differences in the mutant sets.

The mutation analysis follows the structure of the boundary-exploration experiment. Each framework is run 10 times per FUT, and each run produces one archive of discovered boundary candidates. From each archive, we construct one test suite and evaluate it against the fixed mutant set for that FUT. Therefore, for each FUT and framework, we obtain 10 mutation-analysis results, one for each boundary-exploration run, which are later averaged to account for stochastic variation.

Since boundary candidates are input pairs, each candidate contributes two test cases. For numeric FUTs, where both ABEX and SETBVE are applicable, we balance the comparison within each run by first counting the found boundary candidates for each framework and then selecting the smaller of the two counts from both frameworks. This ensures that ABEX and SETBVE are compared using equally sized test suites without giving either framework an advantage due to test-suite size. When one framework discovers more boundary candidates than the selected suite size N, we sub-sample its candidates by sorting them in descending order of program derivative and taking the top N. Since higher PD indicates a stronger behavioral transition and thus a higher likelihood of representing a genuine boundary, this selection favors the framework with the surplus, typically SETBVE, by including its strongest available candidates rather than a random subset. To test whether SETBVE can compensate for lower per-candidate quality with larger test suites, we additionally evaluate a SETBVE-2x configuration that uses up to twice as many SETBVE boundary candidates as ABEX, selected by the same top-PD rule; for FUTs where SETBVE discovers fewer than 2N candidates, all available candidates are used. For non-numeric FUTs, SETBVE is not applicable. Therefore, we report mutation testing results only for ABEX, using the number of boundary candidates available to ABEX. 

For each run-specific test suite, every selected test case is executed on the original program and on each mutant in the fixed mutant set. The results are stored in a kill matrix, where each entry indicates whether a particular test case killed a particular mutant. We compute mutation scores and kill counts separately for each run-specific test suite and then average the results across the runs.

In addition to mutation scores, we analyze whether the generated test suites can detect stubborn mutants, i.e., mutants that are relatively difficult to kill. We use the Relative Stubbornness Thresholding Model (RSTM)~\cite{elgendyeffective}. For each FUT, let $s_i$ be the number of test cases from both methods (ABEX and SETBVE), pooled across all runs, that kill mutant $m_i$, and let $s_{\max}$ be the highest such kill count among all mutants of that FUT. Pooling across both methods ensures a neutral definition of stubbornness: a mutant is classified as stubborn based on its inherent difficulty, not on the limitations of any single method. A mutant is considered stubborn if $s_i \leq \theta \cdot s_{\max}$, where $\theta \in (0,1]$ is a configurable threshold. We use $\theta = 0.05$, meaning a mutant is stubborn if it is killed by at most 5\% of the number of test cases that kill the easiest-to-kill mutant. Once stubborn mutants are identified, we compare how many each method kills (averaged across runs). This defines stubbornness relative to each FUT's own kill distribution instead of relying on a fixed global threshold.

Following standard practice in empirical studies of mutation testing \cite{papadakis2019mutation}, we do not attempt to identify or remove equivalent mutants: reliable detection of equivalent mutants remains an open problem, and manual inspection does not scale to our corpus of 1158 mutants across 20 FUTs. Because identical mutant sets are used for all configurations of the same FUT, any equivalent mutants lower absolute mutation scores for all methods equally and do not bias the relative comparison.

\paragraph{Statistical tests} To assess whether observed differences between configurations are due to chance, we compare per-run outcomes (QD-scores for RQ1, mutation scores for RQ4) using the two-sided Mann–Whitney U test, which makes no assumptions about the underlying distributions \cite{arcuri2014hitchhiker}. We report the Vargha–Delaney $\hat{A}_{12}$ effect size \cite{vargha2000critique}, where $\hat{A}_{12} = 1.00$ indicates that every run of one configuration outperforms every run of the other. Given the small sample size (10 runs per configuration), we use the exact distribution of the U statistic, and we control for multiple comparisons across the FUTs within each pairwise comparison using Holm–Bonferroni correction at $\alpha = 0.05$. FUTs where both configurations produce identical results in all runs are reported as ties and excluded from testing.

\section{Results} \label{sec:results}
This section presents the results of our empirical evaluation.
\subsection{RQ1: Numeric Boundary Discovery} \label{subsec:rq1}

As summarized in Table~\ref{tab:rq1_results}, ABEX achieves the highest QD-score on 10 of 11 FUTs, with an average of 18.7, roughly $2.8\times$ that of single-prompt (6.7) and $11.7\times$ that of SETBVE (1.6), indicating that LLM-generated strategies are effective at discovering boundary candidates that combine high program derivative values with good behavioral diversity. The sole exception is \texttt{bytecount}, where SETBVE achieves a higher QD-score (15.1 vs.\ 10.6), which indicates that traditional search can still be advantageous for some simple numeric structures. 

These differences are statistically significant: ABEX's advantage over SETBVE holds at Holm-corrected $p<0.05$ with maximal effect sizes ($\hat{A}_{12}=1.00$) on 10 of 11 FUTs, and over single-prompt on 10 of 11 FUTs ($\hat{A}_{12} \geq 0.94$). The exceptions are \texttt{bytecount}, where SETBVE's higher QD-score is significant ($p=0.001$, $\hat{A}_{12}=0.09$), and \texttt{complexCheck\_full}, where the difference between ABEX and single-prompt is not significant, consistent with the convergence discussed below.

\begin{table}
\centering
\renewcommand{\arraystretch}{1.2}
\footnotesize
\caption{QD-score and BC count (mean $\pm$ standard deviation over 10 runs) on numeric FUTs. ABEX runs 100 iterations (1000 input pairs), SETBVE runs 1000 iterations (1000 input pairs), and single-prompt issues one call generating 50 input pairs per run. Best QD-score and BC count per FUT in bold.}
\label{tab:rq1_results}
\setlength{\tabcolsep}{1pt}
\begin{tabularx}{\columnwidth}{l YYY YYY}
\toprule
& \multicolumn{3}{c}{\textbf{QD-score}} & \multicolumn{3}{c}{\textbf{BC count}} \\
\cmidrule(lr){2-4} \cmidrule(lr){5-7}
\textbf{FUT} & \textbf{ABEX} & \textbf{Single-Prompt} & \textbf{SETBVE} & \textbf{ABEX} & \textbf{Single-Prompt} & \textbf{SETBVE} \\
\midrule
bmi             & $\mathbf{13.7 \pm 3.9}$ & $6.3 \pm 1.9$ & $0.5 \pm 0.3$ & $65 \pm 21$ & $7 \pm 2$ & $\mathbf{309 \pm 23}$ \\
bytecount       & $10.6 \pm 2.2$ & $6.8 \pm 1.0$ & $\mathbf{15.1 \pm 2.5}$ & $67 \pm 17$ & $8 \pm 2$ & $\mathbf{228 \pm 7}$ \\
calDate         & $\mathbf{19.3 \pm 6.0}$ & $3.1 \pm 0.9$ & $0.3 \pm 0.2$ & $84 \pm 30$ & $15 \pm 2$ & $\mathbf{898 \pm 11}$ \\
circle          & $\mathbf{8.3 \pm 0.7}$ & $2.7 \pm 1.2$ & $0.1 \pm 0.1$ & $32 \pm 10$ & $3 \pm 1$ & $\mathbf{251 \pm 31}$ \\
complexCheck    & $\mathbf{3.0 \pm 1.7}$ & $0.0 \pm 0.0$ & $0.0 \pm 0.0$ & $34 \pm 16$ & $0 \pm 0$ & $\mathbf{235 \pm 17}$ \\
complexCheck\_full & $\mathbf{11.1 \pm 1.6}$ & $10.6 \pm 1.8$ & $0.0 \pm 0.0$ & $58 \pm 27$ & $14 \pm 3$ & $\mathbf{235 \pm 17}$ \\
date            & $\mathbf{25.0 \pm 4.9}$ & $6.6 \pm 2.3$ & $0.0 \pm 0.0$ & $106 \pm 29$ & $19 \pm 2$ & $\mathbf{830 \pm 15}$ \\
english         & $\mathbf{8.7 \pm 1.3}$ & $3.0 \pm 0.6$ & $1.6 \pm 0.9$ & $21 \pm 9$ & $3 \pm 1$ & $\mathbf{30 \pm 8}$ \\
findMiddle      & $\mathbf{41.4 \pm 8.0}$ & $10.4 \pm 2.1$ & $0.2 \pm 0.2$ & $88 \pm 16$ & $18 \pm 4$ & $\mathbf{842 \pm 15}$ \\
nextDate        & $\mathbf{25.3 \pm 7.5}$ & $6.9 \pm 1.5$ & $0.0 \pm 0.0$ & $\mathbf{63 \pm 11}$ & $15 \pm 3$ & $11 \pm 4$ \\
tritype         & $\mathbf{39.0 \pm 6.6}$ & $17.6 \pm 2.6$ & $0.0 \pm 0.0$ & $\mathbf{61 \pm 9}$ & $20 \pm 3$ & $19 \pm 6$ \\
\noalign{\vskip 3pt}
\cdashline{1-7}[0.8pt/2pt]
\noalign{\vskip 3pt}
\textbf{Average} & $\mathbf{18.7}$ & $6.7$ & $1.6$ & $62$ & $11$ & $\mathbf{353}$ \\
\bottomrule
\end{tabularx}
\end{table}

SETBVE dominates in BC count on 9 of 11 FUTs (average 353 vs.\ ABEX's 62), indicating that it is more effective at populating diverse archive regions within the same budget, but at the cost of lower per-cell quality as reflected in its lower QD-scores. This quality gap is substantial: across the 11 numeric FUTs, approximately 99\% of SETBVE's boundary candidates have PD $<$ 0.001, whereas ABEX achieves a mean PD roughly 150$\times$ higher under the same iteration budget. Notably, on \texttt{nextDate} and \texttt{tritype}, ABEX outperforms SETBVE in \textit{both} QD-score and boundary candidate count, indicating that LLM-guided exploration can simultaneously achieve higher coverage and boundary quality for certain FUTs.

The \texttt{complexCheck} pair reveals a sensitivity to specification quality. Without a detailed docstring, single-prompt discovers zero boundary candidates, while ABEX still achieves a QD-score of 3.0 and 34 cells, highlighting its robustness to underspecified functions. With a more detailed specification, both methods improve (ABEX: 11.1; single-prompt: 10.6), and their QD-scores converge, suggesting that when rich context is available, a single prompt can approach the boundary quality achieved by iterative, feedback-driven exploration.

\takeaway[Answer to RQ1]{ABEX achieves higher boundary quality than both baselines on most numeric FUTs, demonstrating that LLM-generated strategies are effective for numeric boundary discovery.}

\subsection{RQ2: Non-numeric Boundaries}

Tables~\ref{tab:qualitative_str} and \ref{tab:qualitative_arr} report the boundary types discovered by ABEX for string and array/mixed FUTs, together with a representative example for each type and the number of runs in which the type was found. Boundary types are derived from each FUT's observable output partitions. For FUTs with discrete semantic outputs, boundary types correspond to transitions between output categories; for example, \texttt{passwordStrength} induces transitions among \{\textit{Weak}, \textit{Acceptable}, \textit{Strong}, \textit{Very strong}\}. For FUTs with structural inputs, boundary types capture input-level transitions that trigger different observable behaviors; for example, \texttt{insertionSort} distinguishes empty from non-empty inputs, changes in array length, changes in element order, and changes in element values.

\begin{table*}
\footnotesize
\renewcommand{\arraystretch}{1.1}
\centering
\caption{Boundary types discovered by ABEX for string-based FUTs, with a representative input pair and the number of runs (out of 10) in which each type was found.}
\label{tab:qualitative_str}
\setlength{\tabcolsep}{6pt}
\begin{tabularx}{\textwidth}{ll>{\raggedright\arraybackslash}Xr}
\toprule
\textbf{FUT} & \textbf{Boundary type} & \textbf{Example: input1 (output1) $\rightarrow$ input2 (output2) } & \textbf{Found runs} \\
\midrule
passwordStrength (str)  & Weak $\leftrightarrow$ Acceptable & onlylowercase (Weak) $\rightarrow$ onlylowercase! (Acceptable) & 10/10 \\
& Acceptable $\leftrightarrow$ Strong & Password123 (Acceptable) $\rightarrow$ Password123! (Strong) & 10/10 \\
& Strong $\leftrightarrow$ Very strong & midlengthnoSym9 (Strong) $\rightarrow$ midlengthnoSym9! (Very strong) & 9/10 \\
& Non-adjacent transition & aaaaaaaaaaaaaaa (Weak) $\rightarrow$ !Aaaaaaaaaaaaaa (Strong) & 9/10 \\
& Invalid $\leftrightarrow$ Weak & [empty] (Invalid) $\rightarrow$ a (Weak) & 10/10 \\

\midrule
printTokens (str) & Token type changed & def (keyword: def) $\rightarrow$ ef (identifier: ef) & 8/10 \\
&Token count changed & 123.456 (number:123.456) $\rightarrow$ 123..456 (number:123.; number:.456) & 10/10 \\ 
 & Token value changed & == (operator: ==) $\rightarrow$ != (operator: !=) & 10/10 \\
 & Valid $\leftrightarrow$ Error & `quote\_here' (string :quote\_here)  $\rightarrow$ `quote\_here (err: unclosed string)& 10/10 \\
\midrule
replace (src, pattern, repl) & Match count changed & `a', `a', `aa' (result:aa; matches:1) $\rightarrow$ `aa', `a', `aa' (result:aaaa; matches:2) & 10/10 \\
  & Same match count, result changed & `a', `a', `A' (result:A; matches:1) $\rightarrow$ `a', `a', `A ' (result:A , matches:1) & 10/10 \\
 & No match $\leftrightarrow$ Match & `A', `a', `b' (result:A; matches:0) $\rightarrow$ `a', `a', `b' (result:b; matches:1) & 10/10 \\

\midrule
stringPalindrome (str) & Palindrome $\leftrightarrow$ Not & 123321 (Palindrome) $\rightarrow$ 1233210 (Not palindrome) & 10/10 \\
\midrule
validateEmail (str) & Valid $\leftrightarrow$ Invalid & bob@trusted.co (Valid) $\rightarrow$ bob@trusted.cö (Invalid format) & 10/10 \\
  & Invalid (different errors) & [empty] (Email must not be empty) $\rightarrow$ [space char] (Invalid format) & 7/10 \\
\bottomrule
\end{tabularx}
\end{table*}

\begin{table*}
\centering
\renewcommand{\arraystretch}{1.1}
\footnotesize
\caption{Boundary types discovered by ABEX for array and mixed input FUTs, with a representative input pair and the number of runs (out of 10) in which each type was found.}
\setlength{\tabcolsep}{6pt}
\label{tab:qualitative_arr}
\begin{tabularx}{\textwidth}{ll>{\raggedright\arraybackslash}Xr}
\toprule
\textbf{FUT} & \textbf{Boundary type} & \textbf{Example: input1 (output1) $\rightarrow$ input2 (output2)} & \textbf{Found runs} \\
\midrule
insertionSort (arr[int]) & Empty $\leftrightarrow$ Non-empty & [] (empty) $\rightarrow$ [0] (single:[0]) & 10/10 \\
& Different array length & [42] (single:[42]) $\rightarrow$ [42,0,0] ([0,0,42]; comp:3; swaps:2) & 10/10 \\
& Same elements different order & [-1,0,0,1,2,3] ([-1,...,3]; comp:5; swaps:0) $\rightarrow$ [-1,0,0,1,3,2] ([-1,...,3]; comp:6; swaps:1) & 10/10 \\
& Different elements & [0] (single:[0]) $\rightarrow$ [1] (single:[1]) & 10/10 \\
\midrule
normalize (arr[int]) & Empty $\leftrightarrow$ Non-empty & [] (Error: Expected...) $\rightarrow$ [0] ([0.0]) & 4/10 \\
  & Different array length & [0.0] ([0.0]) $\rightarrow$ [0.0, 0.0] ([0.0, 0.0]) & 10/10 \\
   & Different elements & [1, 2] ([0.0, 1.0]) $\rightarrow$ [2, 2] ([0.0, 0.0]) & 10/10 \\
\midrule
maxLexString (arr[str]) & Same length, different elements & [`apple', `banana', `cherry'] (cherry) $\rightarrow$ [`apple', `banana', `chery'] (banana) & 10/10 \\
& Single $\leftrightarrow$ Multiple elements & [`file2.txt file10.txt'] (file2.txt file10.txt) $\rightarrow$ [`file2.txt', `file10.txt'] (file10.txt) & 7/10 \\
& Different length, both multiple & [`Z', `z'] (z) $\rightarrow$ [`Z', `z', `zz'] (zz) & 8/10 \\
\midrule
binarySearch (arr[int], int)& Empty $\leftrightarrow$ Non-empty & ([], 1) (empty) $\rightarrow$ ([0], 1) (not\_found) & 10/10 \\
 & Found (diff index) & ([7, 8, 9, 9, 9,...], 9) (found:2) $\rightarrow$ ([7, 8, 8, 9, 9...], 9) (found:3) & 10/10 \\
 & Found $\leftrightarrow$ Not found & ([1, 2, 3, 4, 5,...], 8) (not\_found) $\rightarrow$ ([1, 2, 3, 4, 5,...8...], 8) (found:7) & 10/10 \\
 & Sorted (Valid) $\leftrightarrow$ Unsorted (Error) & ([-3, -2, -1, 0,...], 0) (found:3) $\rightarrow$ ([-3, -2, -1, 5,...], 0) (error:not\_sorted) & 5/10 \\
\midrule
tcas (int, bool, bool, int,...) & Valid $\leftrightarrow$ IndexError & 1024, False, False, 2... (Unresolved) $\rightarrow$ 1024, True, False, 2,... (IndexError) & 4/10 \\
 & Climb $\leftrightarrow$ Unresolved & 601, True, False, 3000... (Climb) $\rightarrow$ 600, True, False, 3000... (Unresolved) & 10/10 \\
 & Descend $\leftrightarrow$ Unresolved &  640, True, False,..., 1,... (Descend)  $\rightarrow$ 640, True, False..., 3,... (Unresolved) & 6/10 \\
\bottomrule
\end{tabularx}
\end{table*}

Once these categories are defined, assigning a discovered boundary candidate to a boundary type is deterministic and rule-based. We do not claim that the listed boundary types form an exhaustive taxonomy of all possible boundaries for each FUT. Different granularities are possible: for example, one could split a broad type into finer subtypes or merge several related transitions
into a coarser category. We use these boundary types as a reproducible summarization of the main behavioral transitions observed in the discovered candidates.

We additionally characterize boundaries using the validity groups from Dobslaw et al.~\cite{dobslaw2023automated}: \textit{VV}, where both outputs are valid but different; \textit{VE}, where one output is valid and the other is erroneous; and \textit{EE}, where both outputs are erroneous but correspond to different errors. When domain-specific output semantics are unavailable, these validity groups provide a coarser fully automatic fallback characterization, because they are computed directly from ABEX's execution outcomes and require no domain-specific boundary-type rules.

\emph{String FUTs.} ABEX consistently discovers meaningful behavioral transitions aligned with domain logic across all string-based programs. For \texttt{passwordStrength}, all five boundary types, including all adjacent strength-level transitions and non-adjacent jumps, are found in at least 9 of 10 runs, demonstrating reliable coverage of a multi-class output space. For \texttt{printTokens}, \texttt{replace}, \texttt{stringPalindrome}, and \texttt{validateEmail}, nearly all boundary types are discovered in 8--10 runs, with the exception of cross-error-type boundaries in \texttt{validateEmail} (7/10). Most discovered boundaries are VV (e.g., crossing password strength thresholds, toggling match counts, breaking palindrome symmetry), while VE boundaries, where one output is erroneous (e.g., unclosed string token, invalid email format), are also discovered consistently, indicating that ABEX is not exclusively biased toward same-validity transitions.

\emph{Array and Mixed FUTs.} For array-based FUTs, ABEX reliably discovers structural boundaries such as empty-vs-non-empty and length-change transitions, with most types found in 8--10 runs. Notable exceptions include the empty-vs-non-empty boundary in \texttt{normalize} (4/10) and the unsorted-array error boundary in \texttt{binarySearch} (5/10), suggesting that VE-type conditions are harder to discover consistently. For the mixed-type \texttt{tcas} FUT, ABEX discovers all three boundary types, including a VE boundary (IndexError) in 4/10 runs, demonstrating that LLM-guided exploration can navigate complex mixed-type input spaces, albeit with reduced consistency on rarer error conditions. 

While the current configuration appears biased toward VV boundaries and may under-emphasize VE and EE transitions, the observed candidates suggest that ABEX can systematically identify rich and interpretable non-numeric boundaries, providing evidence that it generalizes beyond purely numeric domains.

\takeaway[Answer to RQ2]{ABEX consistently discovers meaningful, domain-aligned boundaries across FUTs with string, array, and mixed inputs, demonstrating generalization beyond numeric domains without input-type-specific engineering.}

\subsection{RQ3: Strategy Effectiveness by Input Type} \label{sec:rq3}

Table~\ref{tab:strategy_by_group} reports the effectiveness of each strategy type broken down by FUT group, aggregating results across all 10 runs per FUT\footnote{Strategies discovered per FUT are available in the replication package \texttt{archive/ABEX\_full/strategies}.}. We distinguish four categories: mutation, context-aware generation, generic generation, and the pre-seeded baseline strategy, which is reported separately from context-aware generation as explained in Section \ref{sec:coordinator}. For each strategy type, we report the total number of iterations in which it was selected (\textit{Iterations}), the average number of newly discovered archive cells per 10 generated input pairs (\textit{NC/10}), and the average number of better solutions found for already-occupied cells per 10 input pairs (\textit{BS/10}). NC/10 reflects a strategy's ability to broaden coverage of the behavioral space, while BS/10 reflects its ability to refine the quality of already-discovered archive regions. 

\begin{table}
\footnotesize
\renewcommand{\arraystretch}{1.2}
\centering
\caption{Strategy usage and effectiveness by FUT group: iterations selected, new cells per 10 generated pairs (NC/10), and better solutions in existing cells per 10 pairs (BS/10), aggregated over 10 runs per FUT.}
\label{tab:strategy_by_group}
\begin{tabularx}{\columnwidth}{l l Y Y Y}
\toprule
\textbf{FUT Group} & \textbf{Strategy type} & \textbf{Iterations} & \textbf{NC/10} & \textbf{BS/10} \\
\midrule
Numeric (11 FUTs) & mutation & 3321 & 1.52 & 0.42 \\
& gen/context-aware & 3272 & 0.71 & 0.16 \\
& gen/generic & 2282 & 2.49 & 0.19 \\
& gen/baseline & 2125 & 1.53 & 0.17 \\
\midrule
String (5 FUTs)  & gen/context-aware & 2454 & 2.56 & 0.14 \\
& gen/baseline & 1204 & 3.14 & 0.12 \\
& mutation & 946 & 3.33 & 0.23 \\
& gen/generic & 396 & 3.35 & 0.03 \\
\midrule
Array \& Mixed (5 FUTs) & gen/context-aware & 2089 & 1.59 & 0.36 \\
& gen/baseline & 1395 & 2.28 & 0.27 \\
& gen/generic & 825 & 3.09 & 0.14 \\
& mutation & 691 & 2.70 & 0.39 \\
\bottomrule
\end{tabularx}
\end{table}

\emph{Numeric FUTs.} Mutation is the most frequently selected strategy (3321 iterations), closely followed by context-aware generation (3272), while generic and baseline generation are selected less often (2282 and 2125). Despite its high selection rate, context-aware generation achieves the lowest NC/10 (0.71) and BS/10 (0.16) of all strategies. Generic generation leads in NC/10 (2.49), making it the strongest contributor to coverage breadth, while mutation leads in BS/10 (0.42), outperforming all generation strategies by more than a factor of two. These results reveal a clear division of labor: generic generation drives coverage breadth, while mutation is the dominant mechanism for refining boundary quality.

\emph{String FUTs.} The usage distribution shifts for string FUTs: context-aware generation dominates (2454 iterations), while generic generation is the least used (396). All four strategy types achieve higher NC/10 values than in the numeric group, suggesting the string behavioral space is broadly more amenable to diverse coverage. Mutation achieves the highest BS/10 (0.23), whereas generic generation, despite its leading NC/10 (3.35), contributes negligibly to refinement (0.03). The strong coverage performance of generic generation type indicates that simple string heuristics already capture relevant boundary patterns and that docstrings are not strictly necessary for effective exploration.

\emph{Array and Mixed FUTs.} Context-aware generation is again the most-used strategy type (2089 iterations), while mutation is the least used (691). Generic generation leads in NC/10 (3.09), followed by mutation (2.70) and baseline generation (2.28), mirroring the numeric group's relative rankings. For refinement, mutation and context-aware generation lead with BS/10 of 0.39 and 0.36, respectively. The notably stronger BS/10 of context-aware generation here, compared to numeric and string FUTs, suggests that context captured in docstrings provides more actionable guidance when inputs are compositionally complex.

Overall, generation strategies, especially generic and baseline variants, are critical for discovering diverse archive regions, while mutation is essential for quality refinement. However, none of these strategies operates in isolation during ABEX runs, and it is the coordinated mix of generation and mutation that ultimately yields strong boundary discovery performance.

\takeaway[Answer to RQ3]{Generation strategies drive coverage breadth while mutation refines boundary quality, forming a consistent division of labor whose coordinated interplay is key to boundary discovery performance.}

\subsection{RQ4: Fault Detection}
Table~\ref{tab:mutation-results} reports the mutation testing results averaged across 10 runs. Overall, the results show that boundary candidates discovered by both BVE frameworks are effective at detecting faults. 

\begin{table}[h]
\footnotesize
\renewcommand{\arraystretch}{1.2}
\centering

\caption{Killed mutants and mutation scores (mean $\pm$ standard deviation over 10 runs) for equally sized test suites from ABEX and SETBVE, and for SETBVE-2x with up to twice as many candidates. SETBVE is not applicable to non-numeric FUTs (bottom block).}
\label{tab:mutation-results}

\begin{tabularx}{\textwidth}{
>{\raggedright\arraybackslash}p{1.8cm}
>{\centering\arraybackslash}p{0.8cm}
>{\centering\arraybackslash}p{0.8cm}
>{\centering\arraybackslash}p{1.3cm}
*{5}{>{\centering\arraybackslash}X}
}
\toprule
\textbf{FUT} & \textbf{Tests (avg)} & \textbf{Mutants} &
\multicolumn{3}{c}{\textbf{Killed}} &
\multicolumn{3}{c}{\textbf{Mutation Score (\%)}} \\
\cmidrule(lr){4-6} \cmidrule(lr){7-9}
 & & &
\textbf{ABEX} & \textbf{SETBVE} & \textbf{SETBVE-2x} &
\textbf{ABEX} & \textbf{SETBVE} & \textbf{SETBVE-2x} \\
\midrule
bmi & 130 & 50 &
47.5 $\pm$ 1.3 & 34.8 $\pm$ 4.9 & 35.6 $\pm$ 5.0 &
\textbf{95.0 $\pm$ 2.6} & 69.6 $\pm$ 9.7 & 71.2 $\pm$ 10.0 \\

bytecount & 134 & 65 &
42.5 $\pm$ 1.7 & 30.0 $\pm$ 0.0 & 35.5 $\pm$ 1.9 &
\textbf{65.4 $\pm$ 2.7} & 46.2 $\pm$ 0.0 & 54.6 $\pm$ 2.9 \\

calDate & 168 & 61 &
58.6 $\pm$ 2.5 & 53.2 $\pm$ 1.0 & 53.9 $\pm$ 0.7 &
\textbf{96.1 $\pm$ 4.0} & 87.2 $\pm$ 1.6 & 88.4 $\pm$ 1.1 \\

circle & 63 & 28 &
26.0 $\pm$ 0.0 & 20.2 $\pm$ 2.5 & 21.3 $\pm$ 2.2 &
\textbf{92.9 $\pm$ 0.0} & 72.1 $\pm$ 9.0 & 76.1 $\pm$ 8.0 \\

complexCheck & 116 & 38 &
34.3 $\pm$ 1.3 & 14.4 $\pm$ 0.9 & 14.7 $\pm$ 1.2 &
\textbf{90.3 $\pm$ 3.5} & 37.9 $\pm$ 2.4 & 38.7 $\pm$ 3.1 \\

date & 213 & 7 &
7.0 $\pm$ 0.0 & 7.0 $\pm$ 0.0 & 7.0 $\pm$ 0.0 &
\textbf{100.0 $\pm$ 0.0} & \textbf{100.0 $\pm$ 0.0} & \textbf{100.0 $\pm$ 0.0} \\

english & 37 & 37 &
29.6 $\pm$ 1.2 & 12.5 $\pm$ 1.7 & 13.3 $\pm$ 2.2$^{\dagger}$ &
\textbf{80.0 $\pm$ 3.2} & 33.8 $\pm$ 4.7 & 35.9 $\pm$ 5.9$^{\dagger}$ \\

findMiddle & 176 & 2 &
2.0 $\pm$ 0.0 & 2.0 $\pm$ 0.0 & 2.0 $\pm$ 0.0 &
\textbf{100.0 $\pm$ 0.0} & \textbf{100.0 $\pm$ 0.0} & \textbf{100.0 $\pm$ 0.0} \\

nextDate & 22 & 113 &
71.9 $\pm$ 14.4 & 25.5 $\pm$ 1.7 & 25.5 $\pm$ 1.7$^{\dagger}$ &
\textbf{63.6 $\pm$ 12.8} & 22.6 $\pm$ 1.5 & 22.6 $\pm$ 1.5$^{\dagger}$ \\

tritype & 38 & 85 &
66.9 $\pm$ 5.4 & 42.1 $\pm$ 8.3 & 42.1 $\pm$ 8.3$^{\dagger}$ &
\textbf{78.7 $\pm$ 6.3} & 49.5 $\pm$ 9.8 & 49.5 $\pm$ 9.8$^{\dagger}$ \\

\noalign{\vskip 3pt}
\cdashline{1-9}[0.8pt/2pt]
\noalign{\vskip 3pt}
\textbf{Average} & 109.7 & 48.6 &
38.6 & 24.2 & 25.1 &
\textbf{86.2} & 61.9 & 63.7 \\
\midrule

binarySearch & 343 & 58 &
51.2 $\pm$ 1.3 & -- & -- &
88.3 $\pm$ 2.2 & -- & -- \\

insertionSort & 504 & 49 &
46.0 $\pm$ 0.0 & -- & -- &
93.9 $\pm$ 0.0 & -- & -- \\

maxLexString & 359 & 40 &
12.7 $\pm$ 1.42 & -- & -- &
31.8 $\pm$ 3.5 & -- & -- \\

normalize & 572 & 4 &
4.0 $\pm$ 0.0 & -- & -- &
100.0 $\pm$ 0.0 & -- & -- \\

passwordStrength & 289 & 47 &
45.0 $\pm$ 2.2 & -- & -- &
95.7 $\pm$ 4.7 & -- & -- \\

printTokens & 641 & 205 &
156.0 $\pm$ 7.3 & -- & -- &
76.1 $\pm$ 3.5 & -- & -- \\

replace & 588 & 161 &
73.6 $\pm$ 11.2 & -- & -- &
45.7 $\pm$ 7.0 & -- & -- \\

stringPalindrome & 90 & 11 &
11.0 $\pm$ 0.0 & -- & -- &
100.0 $\pm$ 0.0 & -- & -- \\

tcas & 39 & 71 &
38.0 $\pm$ 5.2 & -- & -- &
53.5 $\pm$ 7.3 & -- & -- \\

validateEmail & 236 & 26 &
20.2 $\pm$ 1.9 & -- & -- &
77.7 $\pm$ 7.5 & -- & -- \\

\noalign{\vskip 3pt}
\cdashline{1-9}[0.8pt/2pt]
\noalign{\vskip 3pt}
\textbf{Average} & 366 & 67.2 &
45.8 & -- & -- &
76.3 & -- & -- \\
 
\bottomrule
\end{tabularx}

\vspace{0.3em}
\footnotesize{$^{\dagger}$ Fewer than twice as many boundary candidates with $PD>0$ were available, so all available candidates were used. These FUTs are included in the SETBVE-2x average.}

\end{table}

When ABEX and SETBVE are compared using test suites of the same size, ABEX achieves higher mutation scores on 8 of the 10 numeric FUTs and ties with SETBVE on the remaining two (\texttt{date} and \texttt{findMiddle}). On average, ABEX reaches a mutation score of 86.2\%, compared to 61.9\% for SETBVE. Excluding the two saturated FUTs, \texttt{date} and \texttt{findMiddle}, which have very few mutants, widens the average mutation-score gap between ABEX and SETBVE. ABEX achieves an average mutation score of 82.8\%, compared with 52.4\% for SETBVE, yielding a difference of about 30. The difference is especially large for \texttt{complexCheck}\footnote{We use the \texttt{complexCheck\_full} variant, which has a more detailed docstring than \texttt{complexCheck}.}, \texttt{english}, \texttt{nextDate}, and \texttt{tritype}, where ABEX improves the mutation score by more than 29 percentage points. These results indicate that, for the same number of tests, ABEX-discovered boundary candidates have stronger fault-detection capability than those discovered by SETBVE. 

These differences are statistically significant with large effect sizes (Mann–Whitney U, Holm-corrected $p<0.05$; $\hat{A}_{12} \geq 0.94$) on all eight numeric FUTs with non-degenerate outcomes; \texttt{date} and \texttt{findMiddle} are ties, with all mutants killed by both methods in every run.

To examine whether SETBVE can close this gap with a larger test suite, we also evaluate SETBVE-2x, using up to twice as many SETBVE boundary candidates as ABEX. This doubled configuration is possible for seven FUTs: \texttt{bmi}, \texttt{bytecount}, \texttt{calDate}, \texttt{circle}, \texttt{complexCheck}, \texttt{date}, and \texttt{findMiddle}. For the remaining three FUTs, \texttt{english}, \texttt{nextDate}, and \texttt{tritype}, SETBVE does not discover enough boundary candidates to construct a doubled suite; in these cases, we use all available SETBVE candidates, marked with $^{\dagger}$ in the table.

SETBVE-2x improves over the SETBVE setting on several FUTs, but the gains are generally small. Its average mutation score (computed over all 10 FUTs, including the four $^{\dagger}$ cases) increases from 61.9\% to 63.7\%, and it still remains below ABEX on 8 of the 10 comparable numeric FUTs, tying only on \texttt{date} and \texttt{findMiddle}. This suggests that the gap is not primarily caused by test-suite size. Even with more boundary candidates, SETBVE does not reach the fault-detection effectiveness of ABEX, indicating that ABEX tends to discover candidates that are more fault-revealing rather than merely more numerous.

The FUTs with non-numeric inputs further show that ABEX-discovered candidates are useful beyond the numeric domain. ABEX achieves high mutation scores for several string, array, and mixed-input functions, including \texttt{passwordStrength} (95.7\%), \texttt{insertionSort} (93.9\%), \texttt{binarySearch} (88.3\%), \texttt{stringPalindrome} (100.0\%), and \texttt{normalize} (100.0\%). The lower scores for \texttt{maxLexString}, \texttt{replace}, and \texttt{tcas} suggest that some non-numeric or mixed-input programs contain faults that are harder to expose with the discovered boundary candidates.

\begin{figure}
    \centering
    \includegraphics[width=0.9\linewidth]{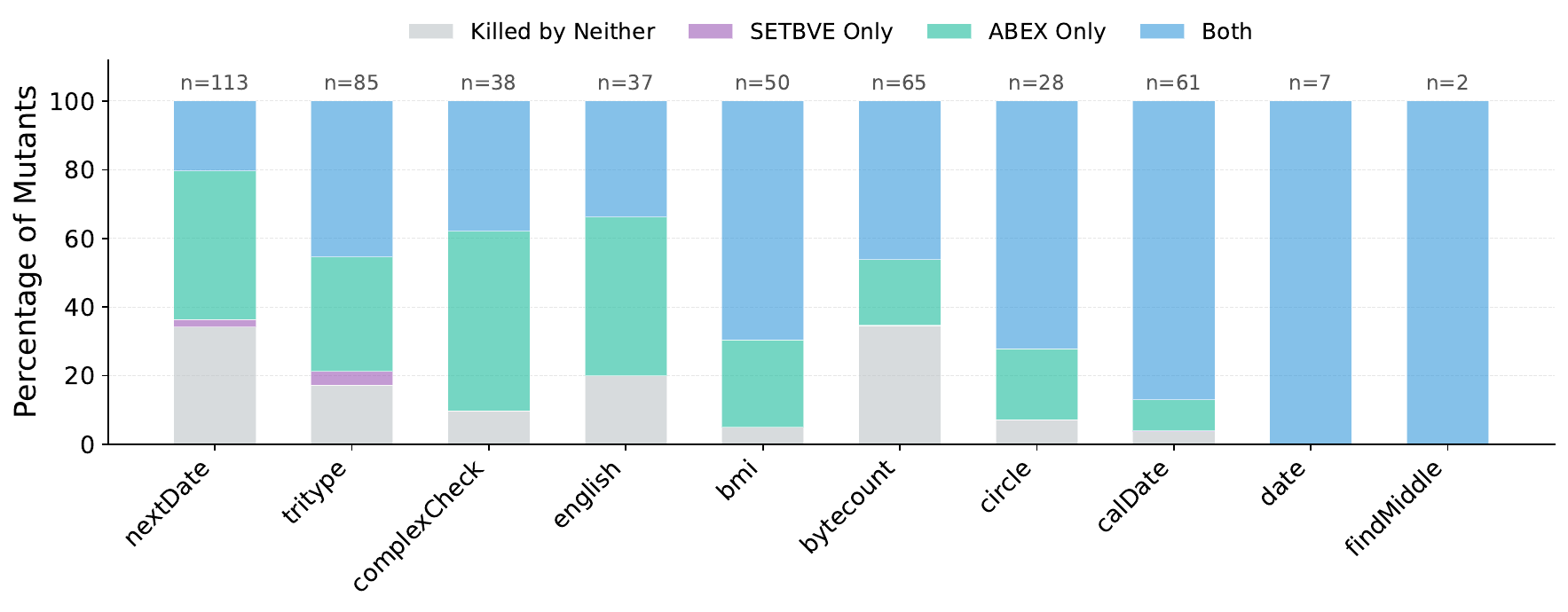}
    \caption{Distribution of mutants killed by ABEX only, SETBVE only, both, or neither, for numeric FUTs (averaged over 10 runs, equal test suite sizes). n gives the number of mutants per FUT.}
    \label{fig:kill_comparison_stacked}
\end{figure}

Figure~\ref{fig:kill_comparison_stacked} further breaks down mutants according to whether they are killed by neither method, only SETBVE, only ABEX, or both methods. The two methods kill many of the same mutants: the overlap is approximately 70\% for \texttt{bmi}, 72\% for \texttt{circle}, 87\% for \texttt{calDate}, and 100\% for both \texttt{date} and \texttt{findMiddle}. The overlap for \texttt{date} and \texttt{findMiddle} is likely due to their small mutant sets, with only 7 and 2 mutants, respectively, reflecting their small code size. ABEX kills many mutants that SETBVE does not. The ABEX-only fraction is largest for \texttt{complexCheck} and \texttt{english}, at roughly 50\% and 46\% of mutants, respectively, followed by \texttt{nextDate} at about 44\% and \texttt{tritype} at about 33\%. In contrast, the SETBVE-only fraction is very small, appearing only for \texttt{nextDate} and \texttt{tritype}.

\begin{figure}
\centering
\begin{minipage}[t]{0.56\textwidth}
\vspace{0pt}
\centering
\includegraphics[width=\linewidth]{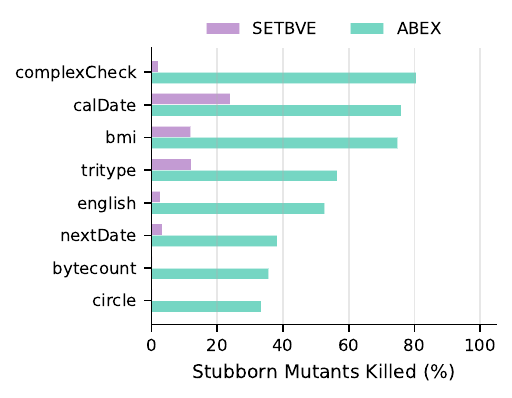}
\caption{Comparison of stubborn mutant detection between ABEX and SETBVE averaged over 10 runs (RSTM $\theta=0.05$).}
\label{fig:stubborn_mutants}
\end{minipage}
\hfill
\begin{minipage}[t]{0.42\textwidth}
\vspace{-20pt}
\centering
\small
\captionof{table}{Number of stubborn mutants killed by ABEX and SETBVE per numeric FUT (mean over 10 runs, RSTM $\theta=0.05$), out of the total per FUT.}
\label{tab:stubborn-mutants}
\renewcommand{\arraystretch}{1.2}
\begin{tabular}{lrrr}
\toprule
\textbf{FUT} & \textbf{Total} & \textbf{SETBVE} & \textbf{ABEX} \\
\midrule
bmi & 10 & 1.2 & 7.5 \\
bytecount & 35 & 0.0 & 12.5 \\
calDate & 10 & 2.4 & 7.6 \\
circle & 3 & 0.0 & 1.0 \\
complexCheck & 19 & 0.4 & 15.3 \\
english & 15 & 0.4 & 7.9 \\
nextDate & 64 & 2.1 & 24.5 \\
tritype & 34 & 4.1 & 19.2 \\
\noalign{\vskip 3pt}
\cdashline{1-4}[0.8pt/2pt]
\noalign{\vskip 3pt}
\textbf{Total} & \textbf{190} & \textbf{10.6} & \textbf{95.5} \\
\bottomrule
\end{tabular}
\end{minipage}

\end{figure}

Figure~\ref{fig:stubborn_mutants} and Table ~\ref{tab:stubborn-mutants} compare the number of stubborn mutants killed by ABEX and SETBVE. Stubborn mutants are those killed by at most 5\% as many test cases as the easiest-to-kill mutant for the same FUT (see Section~\ref{subsec:experimental_setup} for details). ABEX kills more stubborn mutants than SETBVE for every FUT shown\footnote{\texttt{findMiddle} and \texttt{date} are excluded because no stubborn mutants were detected for these FUTs.}. The largest absolute numbers occur for \texttt{nextDate}, \texttt{tritype}, and \texttt{complexCheck}, where ABEX kills, on average over 10 runs, 24.5, 19.2, and 15.3 stubborn mutants, respectively. In contrast, SETBVE kills only a small number of stubborn mutants for these functions and kills almost none for \texttt{complexCheck}, \texttt{english}, and \texttt{bytecount}. Overall, ABEX kills 95.5 of the 190 stubborn mutants (50.3\%), while SETBVE kills 10.6 (5.6\%). Thus, SETBVE kills only about 11\% as many stubborn mutants as ABEX. 

The stubborn-mutant distribution also explains the notably high run-to-run variance of ABEX on \texttt{nextDate} ($\sigma$ = 12.8 percentage points, the highest in Table \ref{tab:mutation-results}). \texttt{nextDate} contains the largest concentration of stubborn mutants in our corpus (64 of 113, 57\%), located primarily in its leap-year and month-boundary logic. Whether these mutants are killed appears to depend on whether a given run discovers strategies targeting those specific edge cases: runs that do reach mutation scores of up to 81\%, while runs that do not fall to 37\%. The other date-related functions lack this concentration, \texttt{date} has no stubborn mutants, and \texttt{calDate} has 16\%, and correspondingly show low variance.

For FUTs with string, array, and mixed input types, no SETBVE baseline exists. This group includes eight functions, excluding \texttt{normalize} and \texttt{stringPalindrome}, for which no stubborn mutants were detected. On these eight functions, ABEX kills an average of 74.9 out of 279 stubborn mutants in total, corresponding to a kill rate of 26.8\%. The kill rate varies across FUTs: ABEX achieves over 35\% on five FUTs (\texttt{passwordStrength}, \texttt{printTokens}, \texttt{insertionSort}, \texttt{binarySearch}, and \texttt{validateEmail}), but performs less effectively on \texttt{maxLexString} and \texttt{tcas}, with kill rates of 2.5\% and 14.2\%, respectively. These two functions contain 28 and 33 stubborn mutants, respectively. Overall, the kill rate for this group of functions is lower than for functions with integer inputs, which may be due to differences in function complexity and greater input variability.

\takeaway[Answer to RQ4]{ABEX-discovered boundary candidates are more fault-revealing than SETBVE's, achieving higher mutation scores and killing more difficult-to-detect stubborn mutants.}
  
\subsection{RQ5: Ablation} \label{subsec:ablation}

We conduct RQ5 on a representative subset of six FUTs selected from the initial set of 20: \texttt{bmi}, \texttt{tritype}, \texttt{passwordStrength}, \texttt{replace}, \texttt{insertionSort}, and \texttt{binarySearch}. These functions were selected to cover variation in input type, output type, arity, and boundary complexity. The subset includes numeric, string, and array-based inputs; categorical, transformed, and index outputs; one-, two-, and three-parameter functions; and different forms of boundary behavior, including threshold-based boundaries, multi-condition classification, pattern matching, structural array behavior, and precondition-dependent behavior.

\begin{figure}
    \centering
    \includegraphics[width=0.7\linewidth]{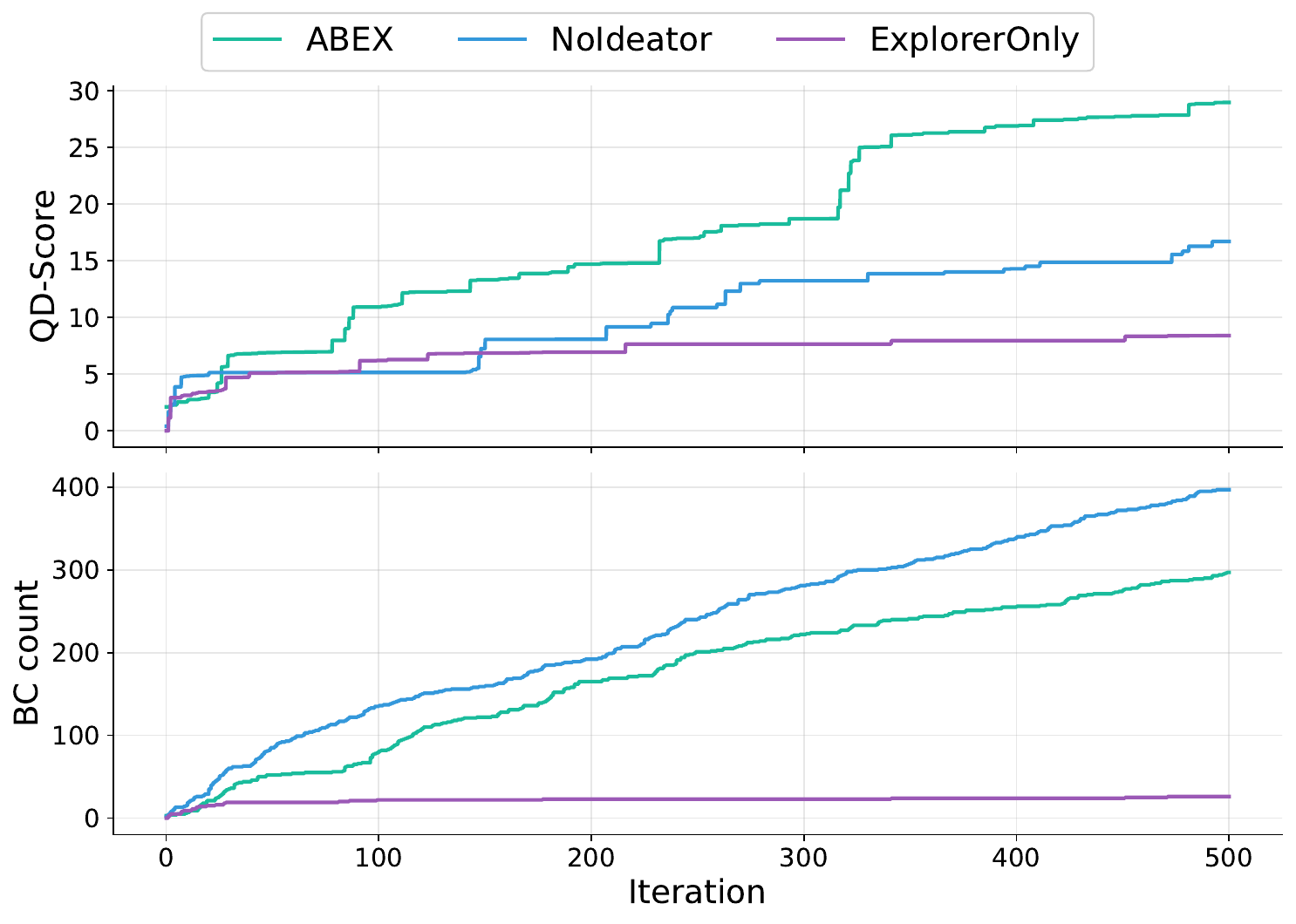}
    \caption{QD-score and BC count trend for the \texttt{bmi} FUT.}
    \label{fig:ablation_line}
\end{figure}

Unlike the RQ1 -- RQ2 experiments, the ablation and LLM-comparison runs use 500 iterations rather than 100. Longer runs are needed here because differences between configurations that share components can take longer to manifest than differences between distinct frameworks: a weaker variant may keep pace early while the archive is sparse, and fall behind only once easy boundary candidates are exhausted (cf. Figure \ref{fig:ablation_line}). Because of the associated cost, each configuration is run once per FUT on the representative subset.

\begin{figure}
    \centering
    \includegraphics[width=0.7\linewidth]{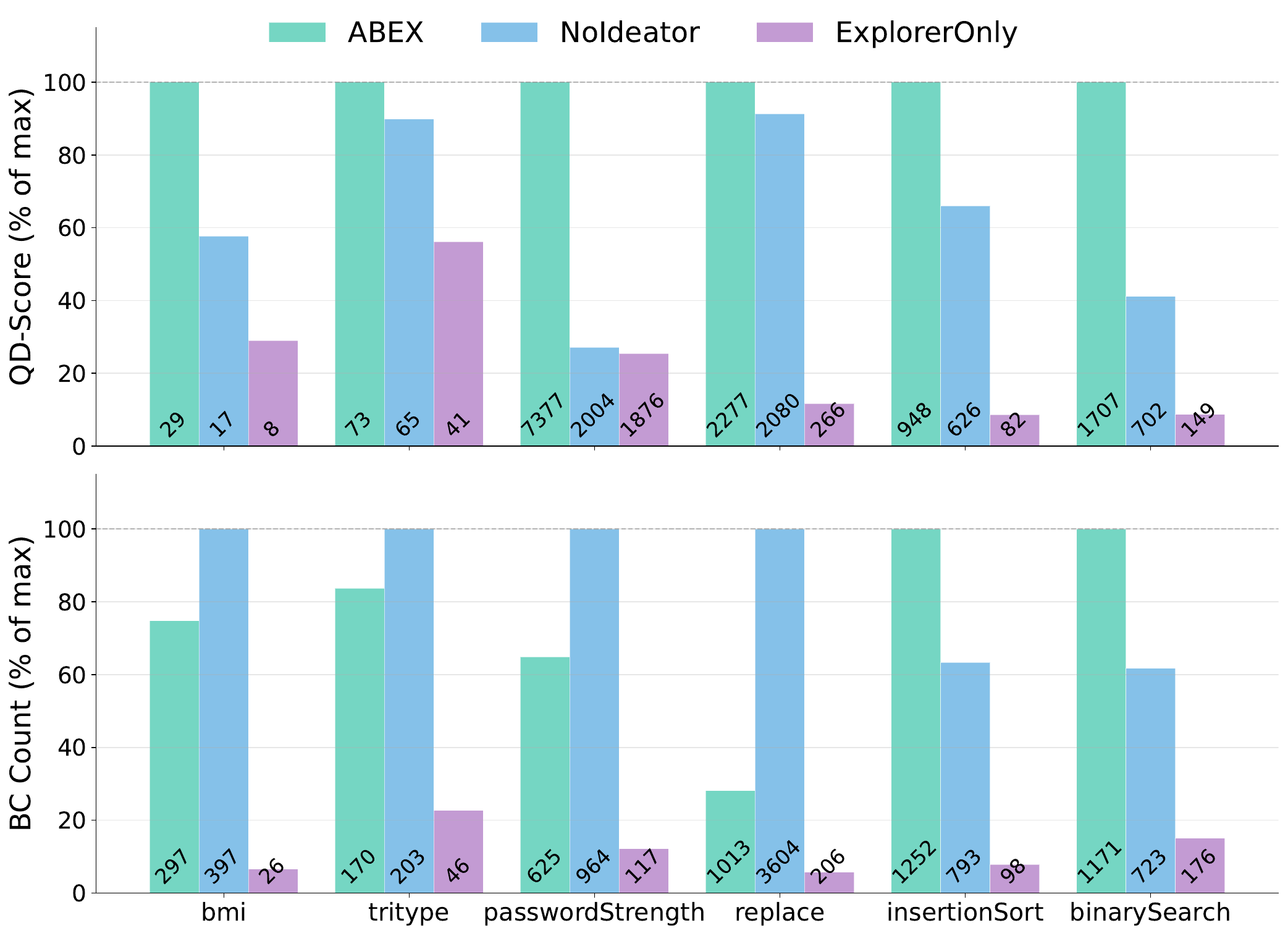}
    \caption{Component ablation on six FUTs: QD-score (top) and BC count (bottom) for ABEX, NoIdeator, and ExplorerOnly, normalized per FUT to the best performing variant (bar labels show absolute values).}
    \label{fig:ablation_bar}
\end{figure}

\paragraph{Components ablation} Figure~\ref{fig:ablation_bar} compares ABEX against two ablated variants. \emph{NoIdeator} removes the Ideator: the Strategy Generator expands the Coordinator's search gap directly into a strategy, without the intermediate step that compares candidate directions against the existing strategy pool for functional novelty, reuses existing strategies when a suitable one is available, and filters directions likely to be low-yield before expansion. \emph{ExplorerOnly} removes the Coordinator, Ideator, and Strategy Generator entirely and executes only the pre-seeded baseline strategy throughout the run; it retains the same archive feedback as the full framework, with each Explorer prompt including recently added and randomly sampled input pairs to avoid duplicates. All three configurations generate the same number of candidate input pairs (10 per iteration over 500 iterations), so differences reflect the removed components rather than generation budget. Figure~\ref{fig:ablation_line} shows QD-score and boundary candidate count over 500 iterations for \texttt{bmi} as a representative example; further plots are in the replication package.

ABEX achieves the highest QD-score on all six FUTs, confirming that the full pipeline is necessary. ExplorerOnly performs worst relative to the ablation variants across all FUTs in both metrics, establishing adaptive strategy generation as the primary driver of effectiveness. Averaged over the six FUTs, ABEX improves QD-score by $6.8\times$ over ExplorerOnly (per-FUT ratios ranging from $1.8\times$ on \texttt{tritype} to $11.6\times$ on \texttt{insertionSort}). The NoIdeator setup falls between the two, showing the Ideator yields measurable quality gains. NoIdeator achieves higher boundary cell counts than ABEX on several FUTs, suggesting that, for those FUTs, it discovers more diverse boundary candidates, but of lower quality. As shown in Figure \ref{fig:ablation_line}, QD-score growth is non-monotonic, with sharp jumps coinciding with strategy switches, visible in ABEX and NoIdeator but absent in ExplorerOnly. Moreover, both ABEX and NoIdeator show no signs of stagnation even after almost 1 hour (500 iterations), suggesting that longer runs could yield further improvements.

\paragraph{Cost} Runtime cost depends on FUT complexity and strategy turnover rate. The framework uses a two-tier model architecture: GPT-5.1 handles the Coordinator and Ideator, while GPT-5-mini powers the Explorer and Strategy Generator. On average, ABEX consumed 2.7M tokens (6.38 USD) per 500-iteration run, with 416K tokens allocated to GPT-5.1 and 2.3M to GPT-5-mini. NoIdeator reduced costs to 2.3M tokens (5.72 USD) by eliminating Ideator calls while retaining strategy generation overhead. ExplorerOnly, lacking both Coordinator and Ideator, required only 1.4M tokens (2.71 USD), representing cost reduction compared to ABEX, albeit with a trade-off in quality.

\paragraph{Closed vs. open-weight LLMs} To evaluate whether the choice of LLM affects ABEX's boundary candidate generation ability, we compare three LLMs: GPT-5.1/mini (closed-weight), Gemma-4 (open-weight), and Qwen-3.5 (open-weight). We run ABEX with each LLM on six functions for 500 iterations, measuring BC Count and QD-score. Note that ABEX-GPT5.1/mini uses a combination of GPT-5.1 and GPT-5-mini to reduce API costs, whereas ABEX-Gemma4 and ABEX-Qwen3.5 use the same full model for all ABEX components throughout.

\begin{figure}
    \centering
    \includegraphics[width=0.7\linewidth]{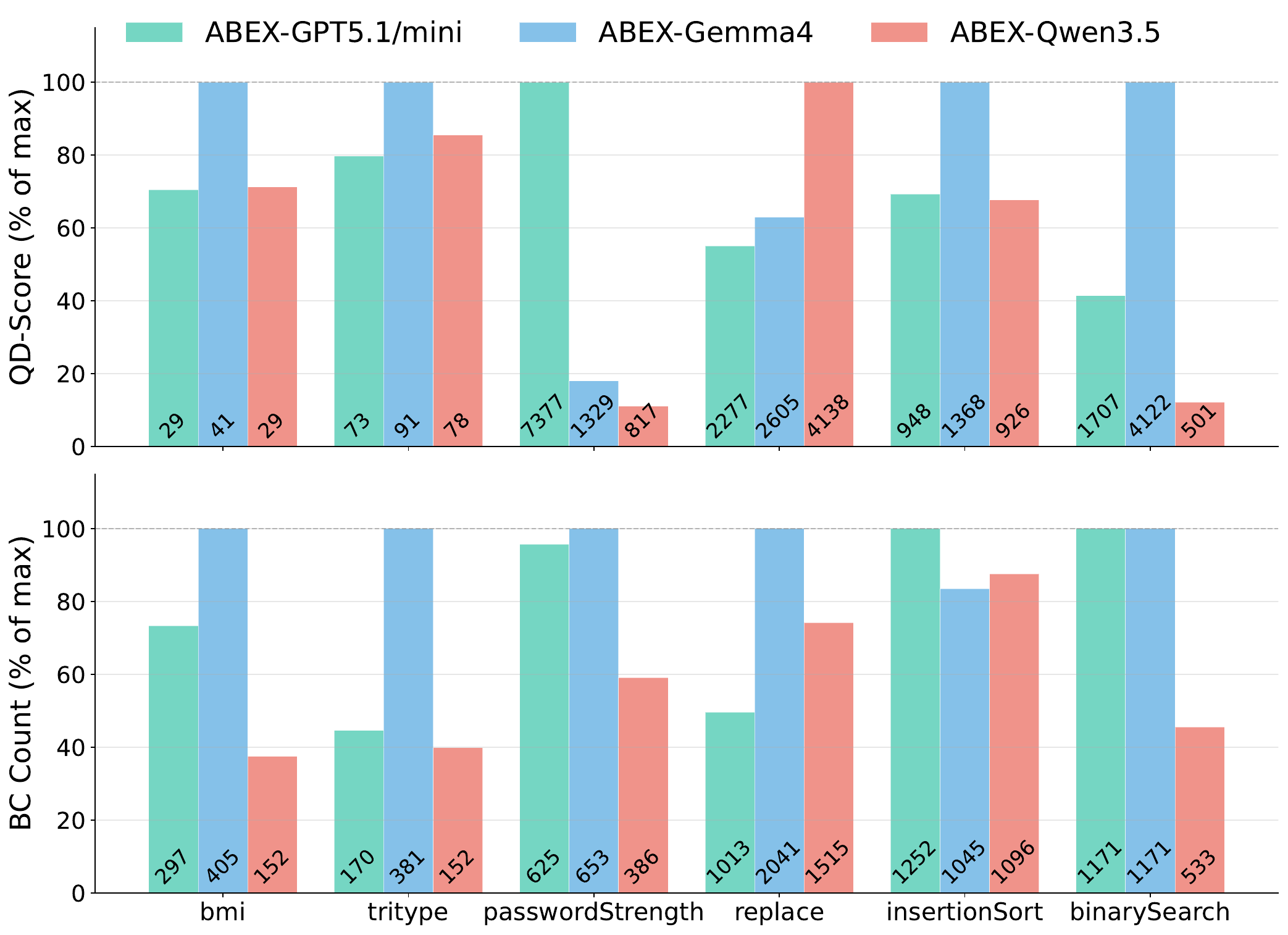}
    \caption{ABEX with three LLMs (GPT-5.1/mini, Gemma-4, Qwen-3.5) on six FUTs: QD-score (top) and BC count (bottom), normalized per FUT (bar labels show absolute values).}
    \label{fig:llm_comparison}
\end{figure}

All three models successfully discovered boundary candidates across all six functions. Figure~\ref{fig:llm_comparison} shows the results. Gemma-4 achieved the highest total BC count (5696), followed by GPT-5.1/mini (4528) and Qwen-3.5 (3834). Gemma-4 achieved the maximum BC count on four of six functions (\texttt{bmi}, \texttt{tritype}, \texttt{passwordStrength}, \texttt{replace}), while GPT-5.1/mini achieved the maximum on two functions (\texttt{insertionSort}, \texttt{binarySearch}, tied with Gemma-4 on the latter).

The QD-Score results provide a complementary view of model performance. Gemma-4 obtains the highest QD-score on four functions\footnote{Gemma-4 was released in April 2026, after the RQ1 -- RQ4 experiments had been completed. The main ABEX configuration therefore uses GPT-5.1/mini.}: \texttt{bmi}, \texttt{tritype}, \texttt{insertionSort}, and \texttt{binarySearch}. This indicates that, for these functions, Gemma-4 not only discovers many candidates but also accumulates high-quality candidates overall. GPT-5.1/mini achieves the highest QD-score on \texttt{passwordStrength}, despite discovering slightly fewer candidates than Gemma-4 for this function. This suggests that GPT-5.1/mini identifies fewer but much stronger candidates in this case. Qwen-3.5 achieves the highest QD-score on \texttt{replace}, where it outperforms both GPT-5.1/mini and Gemma-4 in accumulated quality, even though Gemma-4 discovers the largest number of candidates for this function. 

To further assess candidate quality, we compute the average program derivative (QD-score / BC Count)\footnote{Average PD is computed from raw program-derivative values, whose range is FUT-dependent.} for each model-function combination (Table~\ref{tab:llm-avg-pd}). GPT-5.1/mini achieved the highest mean average PD (2.80), compared to Gemma-4 (1.41) and Qwen-3.5 (1.22). GPT-5.1/mini's advantage is driven primarily by \texttt{passwordStrength}, where it achieved an average PD of 11.80 compared to 2.04 (Gemma-4) and 2.12 (Qwen-3.5).

\begin{table}[h]
\renewcommand{\arraystretch}{1.2}
\small
\centering
\caption{Average program derivative (QD-score / BC Count) by LLM over 500 iterations.}
\label{tab:llm-avg-pd}
\begin{tabular}{lrrr}
\toprule
\textbf{FUT} & \textbf{ABEX-GPT5.1/mini} & \textbf{ABEX-Gemma4} & \textbf{ABEX-Qwen3.5} \\
\midrule
bmi & 0.10 & 0.10 & 0.19 \\
tritype & 0.43 & 0.24 & 0.51 \\
passwordStrength & 11.80 & 2.04 & 2.12 \\
replace & 2.25 & 1.28 & 2.73 \\
insertionSort & 0.76 & 1.31 & 0.84 \\
binarySearch & 1.46 & 3.52 & 0.94 \\
\noalign{\vskip 3pt}
\cdashline{1-4}[0.8pt/2pt]
\noalign{\vskip 3pt}
\textbf{Average} & \textbf{2.80} & \textbf{1.41} & \textbf{1.22} \\
\bottomrule
\end{tabular}
\end{table}

The relationship between quantity and quality varies across functions. For \texttt{replace}, Gemma-4 discovered the most candidates (2041) but achieved the lowest average PD (1.28). For \texttt{binarySearch}, Gemma-4 and GPT-5.1/mini discovered identical BC counts (1171), but Gemma-4 achieved a higher average PD (3.52 vs. 1.46). Qwen-3.5 achieved the highest average PD on three functions (\texttt{bmi}: 0.19, \texttt{tritype}: 0.51, \texttt{replace}: 2.73) despite having lower BC counts.

Both closed and open-weight LLMs generate meaningful boundary candidates with ABEX. Gemma-4 produces the highest boundary candidate count, GPT-5.1/mini produces higher-quality candidates on average despite using a mixed-model configuration, and Qwen-3.5 shows competitive quality on specific functions but lower overall quantity of boundary candidates.

\takeaway[Answer to RQ5]{Adaptive strategy generation is the primary effectiveness driver. The absence of stagnation suggests further improvement is achievable with longer runs. ABEX remains effective across both closed and open-weight LLMs, although the magnitude of performance gains varies across functions and models.}

\section{Discussion} \label{sec:discussion}
ABEX shows that an agentic, LLM-based framework can automate boundary value exploration across heterogeneous input types without manually engineered mutation operators, while producing higher-quality boundaries than both baselines on numeric inputs. On 10 of 11 numeric FUTs, ABEX achieves the highest QD-scores, averaging $2.8\times$ the single-prompt LLM baseline and $11.7\times$ the QD-based search baseline. For non-numeric inputs, explored here for the first time in automated BVE in a black-box setting, ABEX discovers meaningful, domain-aligned boundary behaviors across all 10 string, array, and mixed input FUTs, again without datatype-specific operators. Taken together, these results show that adaptive, LLM-guided search can extend BVE beyond numeric domains.

\paragraph{Adaptive strategy generation matters} The comparison between ABEX and the single-prompt baseline is revealing because both use the same underlying LLM to generate inputs. Single-prompt achieves an average QD-score of 6.7, whereas ABEX reaches 18.7. This gap therefore cannot be explained by model capability alone; it points instead to the agentic design and adaptive feedback loop as the main sources of boundary quality. Because the two configurations use different generation budgets (Section \ref{subsec:experimental_setup}), part of this gap could in principle be attributed to budget rather than design. The ablation results let us separate the two factors. ExplorerOnly iterates the same fixed baseline strategy as single-prompt, with archive feedback, under a full search budget; yet even with more generated candidate pairs, it remains far below the full framework on the shared FUTs (\texttt{bmi}: 8 vs. 29; \texttt{tritype}: 41 vs. 73), with \texttt{bmi} improving only from 6.3 to 8 over the single call. Budget-matched iteration of a fixed strategy thus recovers only part of the difference; adaptive strategy generation accounts for the rest, with NoIdeator's intermediate performance further showing that adaptive strategy selection and Ideator-mediated refinement make distinct contributions to quality. This interpretation aligns with Lemieux et al.~\cite{lemieux2023codamosa}, who show that feedback-driven LLM exploration outperforms one-shot generation, and with the multi-agent testing results of Yoon et al.~\cite{yoon2024intent}.

The strategy-level analysis (RQ3) explains \emph{how} adaptation helps: generation strategies broaden archive coverage while mutation strategies refine boundary quality. This pattern mirrors the classic exploration--exploitation trade-off in evolutionary search, so it is not surprising in direction. Its importance here is that it appears uniformly across numeric, string, and array domains, indicating that ABEX benefits from genuine complementarity among strategy types rather than from any one strategy dominating the search.

\paragraph{Complementarity with optimized QD search} ABEX and SETBVE~\cite{akbarova2025setbve} make different trade-offs. Under a fixed iteration budget, SETBVE discovers more boundary candidates than ABEX on average (353 vs.\ 62 BC count), confirming that optimized numeric search is more effective at broadly covering the behavioral space. ABEX, by contrast, prioritizes boundary quality. As reported in Section~\ref{subsec:rq1}, approximately 99\% of SETBVE's candidates have $\mathrm{PD} < 0.001$, whereas ABEX achieves a mean PD roughly 150$\times$ higher under the same budget. Since the BC count only requires $\mathrm{PD} > 0$, many near-zero-PD boundary candidates may reflect different equivalence partitions rather than true boundary cases.
 
The trade-off between quality and breadth is, moreover, not universal: on \texttt{nextDate} and \texttt{tritype}, ABEX outperforms SETBVE on both archive coverage and QD-score. The main exception in the other direction is \texttt{bytecount}, where SETBVE achieves a higher QD-score (15.1 vs.\ 10.6), suggesting that numerically simple functions with tightly bounded input structures remain well suited to optimized search. Overall, these results position ABEX as a complement to QD-based search rather than a replacement: ABEX provides higher
boundary quality and supports non-numeric inputs, but at a higher compute cost per iteration. For numeric inputs, a natural hybrid strategy would use SETBVE for broad exploration and ABEX to refine promising regions.

\paragraph{The role of specification quality} The \texttt{complexCheck} results reveal a robustness property with clear practical value. With only a short docstring, ABEX achieves a QD-score of 3.0 and discovers 34 boundary cells, whereas single-prompt produces no candidates at all. With a more detailed specification, both methods improve and their QD-scores converge (11.1 vs.\ 10.6), suggesting that richer context narrows the advantage of iteration but does not eliminate it. A natural next step is to relax the black-box assumption progressively and test whether richer context yields proportionate gains. The non-numeric results also show a tendency toward VV boundaries, which likely reflects the default prompting configuration. Because strategy generation is LLM-driven, explicitly targeting VE or EE validity groups through tailored prompts is a straightforward next step that requires no architectural changes.

\paragraph{Fault-detection effectiveness} Mutation testing confirms that ABEX-discovered boundary candidates are effective as fault-detecting test cases. Compared with SETBVE under equal test-suite sizes, ABEX achieves higher mutation scores on 8 of 10 numeric FUTs, and SETBVE does not close the gap even when given up to twice as many boundary candidates (SETBVE-2x). The gap is therefore not primarily an artifact of test-suite size: ABEX tends to discover candidates that are more fault-revealing, not merely more numerous. The stubborn-mutant analysis sharpens this point. ABEX kills 50.3\% of stubborn mutants on numeric FUTs versus 5.6\% for SETBVE, indicating that the quality advantage translates into the ability to expose faults.

\paragraph{No dependence on proprietary frontier models} The LLM comparison (RQ5) shows that ABEX's effectiveness is not tied to a specific proprietary model: the open-weight Gemma-4 achieved the highest boundary candidate count and the highest QD-score on four of six FUTs, while GPT-5.1/mini produced higher average per-candidate quality. This matters for practical adoption. Organizations with confidentiality constraints that preclude sending code artifacts, signatures, or specifications to third-party APIs can run ABEX entirely on locally hosted open-weight models, and open-weight deployment also removes per-token API costs from the framework's main cost driver (Section \ref{subsec:ablation}). The performance differences across FUTs suggest that model choice can be treated as a tunable deployment decision rather than a prerequisite for the approach.

\paragraph{Implications for research and practice} The main implication for researchers is that adaptive, LLM-guided search is a viable general mechanism for testing problems in which operator engineering is costly or infeasible. We demonstrate this here for BVE, but the core ingredients should transfer to any setting in which exploration policies can be expressed and assessed through execution feedback. For practitioners, the clearest benefit lies in non-numeric inputs or settings where per-type operator design is substantial. For simple numeric FUTs, SETBVE remains faster and cheaper, though that gap will likely narrow as inference costs fall~\cite{gundlach2025price}.

\subsection{Lessons Learnt and Future Directions}
We draw three main lessons from this study. 

First, behavioral descriptors strongly influence which boundaries the search discovers. We reused the four descriptors from SETBVE because they generalize across input types and worked well overall, but for array-based FUTs they often distinguish small input changes without capturing genuinely different \emph{types} of behavioral transitions. For structured inputs, better FUT-specific descriptors may therefore be needed, potentially with LLM support.

Second, the Coordinator does not always know when the search has run out of useful directions. For simple or tightly constrained FUTs, such as \texttt{circle} or \texttt{complexCheck}, exploration saturates quickly, yet the system may still request new strategies. A \emph{dynamic stopping agent} could improve efficiency and cost predictability by tracking saturation signals, such as QD-score change relative to Coordinator invocation frequency, and then halting, redirecting, or escalating to the tester for guidance.

Third, the strategy library matters beyond initialization. In our runs, it provided the starting point through the baseline strategy, but it can also serve as a reusable knowledge base: effective strategies can transfer across FUTs, and testers can seed the library with domain-specific strategies when prior knowledge exists. We note that while the strategies produced during our runs are human-readable, we did not empirically evaluate their interpretability to testers, nor did we measure the benefit of transferring strategies across FUTs. Both are natural next steps for validating the strategy library as an accumulating knowledge asset.

Taken together, these lessons suggest that ABEX is best viewed as a platform that can improve with use. As the strategy library grows and coordination becomes more selective, exploration knowledge can become a durable asset rather than something recreated in each run.

\subsection{Validity Threats}

\paragraph{Construct validity} We use QD-score as the primary metric, following standard practice in QD evaluation \cite{xiang2023automated}, and complement it with boundary-candidate cell count. SETBVE \cite{akbarova2025setbve} reports RAC and RPD, but these require substantial archive-cell overlap between the compared methods, and overlap in our study is too limited because ABEX and SETBVE explore the space differently. Both QD-score and BC count are, however, proxies for boundary quality and diversity rather than direct measures of fault-detection value. Moreover, the ingredients they are computed from are heuristic choices inherited from prior work \cite{akbarova2025setbve}: the program derivative used as fitness relies on fixed distance metrics rather than ground-truth measures of boundariness. In particular, we compute output distances using Jaccard distance on stringified outputs, so transitions whose outputs share similar n-gram sets may receive low PD and be undervalued. The behavioral descriptors determine what the archive treats as ``diverse''. Alternative fitness functions, output-distance measures, or descriptor sets could therefore lead to different archive structures and potentially shift the results. Since all configurations use the same metrics, these choices affect PD and QD-score magnitudes rather than the fairness of the comparison. To link the discovered candidates to fault detection through a metric-independent lens, we complement the archive-level analysis with mutation testing (RQ4), which confirms that the candidates are fault-revealing. Mutation testing has its own construct limitation: it measures fault detection against syntactic mutants produced by mutmut's operator set, which are proxies for real faults \cite{papadakis2019mutation}, and since we do not filter equivalent mutants, absolute mutation scores should be read comparatively rather than as fault-detection rates for real defects.

\paragraph{Internal validity} We compared ABEX and SETBVE under a fixed budget of generated input pairs rather than wall-clock time. This keeps the comparison aligned at the level of exploration effort, but it also favors neither method's runtime characteristics: SETBVE is faster per iteration because it does not invoke LLMs, so a time-budget comparison would likely improve its raw throughput. At the same time, LLM costs are falling rapidly; Gundlach et al.~\cite{gundlach2025price} estimate declines of 5--10$\times$ per year, with algorithmic efficiency alone contributing about 3$\times$ annually. We therefore report a hardware-agnostic comparison that readers can map to current runtimes and costs. Also, the single-prompt baseline uses a smaller generation budget than ABEX, following prior work \cite{guo2025boundary}. In preliminary experiments, we attempted to scale the single call to larger candidate sets, but the model increasingly produced near-duplicate pairs, so a larger single-call budget does not yield a proportionally stronger baseline. The natural way to scale a single prompt is to invoke it repeatedly, which is precisely the ExplorerOnly configuration (Section \ref{subsec:ablation}): it executes the same fixed strategy with archive feedback under a full search budget, yet its QD-score and BC count saturate early (Figure \ref{fig:ablation_line}), with generated pairs becoming repetitive despite archive-based duplicate avoidance. This indicates that the gap between ABEX and single-prompt is not primarily a budget artifact. Finally, both single-prompt and ExplorerOnly rely on a single fixed baseline strategy; a stronger baseline with several fixed strategies would better isolate the contribution of adaptive strategy selection.

\paragraph{External validity} We evaluate 20 FUTs across four input categories and a range of program complexities. Although broader than prior BVE benchmarks, all FUTs are small, stateless, side-effect-free, single-function Python units with English-language docstrings; the behavior of ABEX on stateful code, multi-function APIs, larger real-world programs, or non-English specifications remains untested. LLM's training data contamination is a potential threat, since several subject programs (e.g., \texttt{tritype}, \texttt{tcas}, \texttt{nextDate}) are long-standing testing benchmarks and may appear in training data. This raises the possibility that some boundary knowledge is memorized rather than inferred from the signature and docstring. However, ABEX and the single-prompt baseline use the same model, so memorization alone cannot explain the 2.8$\times$ QD-score gap; moreover, the black-box protocol reveals only the signature and docstring, not the implementation. Evaluation on proprietary or newly written functions is still needed to fully rule out contamination.

Results may also vary across LLM models and configurations, an underexplored source of instability in LLM-based systems \cite{dobslaw2025challenges}. To partially assess this threat, we evaluated ABEX with three different LLMs: GPT-5.1/mini as a closed-weight proprietary model, and Gemma-4 and Qwen-3.5 as open-weight models. In all cases, ABEX was able to generate boundary exploration strategies and identify boundary candidates. However, performance varied across FUTs, suggesting that different LLMs may be better suited to different types of functions. At minimum, ABEX requires an LLM that can reason about boundary testing and input data types well enough to generate meaningful input mutations. Future work should investigate which LLM characteristics are most effective for different categories of FUTs.

\paragraph{Conclusion validity} The main threat here is stochastic variation. We repeated each configuration 10 times per FUT and assessed the main comparisons with non-parametric significance tests and effect sizes. The consistently large $\hat{A}_{12}$ values suggest the reported differences are robust to run-to-run variation. Nevertheless, 10 runs may be too few to characterize variance fully for string, array, and mixed FUTs, where output variability is higher and where no baseline comparison was tested. The ablation and LLM-comparison experiments use a single 500-iteration run per configuration due to cost; their results should therefore be read as indicative.

\section{Conclusion} \label{sec:conclusion}

We presented ABEX, an agentic LLM framework for automated boundary value exploration that eliminates the need for manually engineered mutation operators and extends BVE beyond numeric inputs. By combining quality-diversity search with adaptive, LLM-driven strategy generation, ABEX enables boundary discovery across heterogeneous input types within a single unified framework. Our results show that ABEX consistently produces higher-quality boundary candidates than both traditional search-based methods and a baseline LLM approach. In numeric domains, it achieves the highest QD-scores on most FUTs, while for non-numeric inputs it discovers meaningful, domain-aligned boundary behaviors without requiring datatype-specific operators. Mutation testing confirms that these boundary candidates translate into fault-detection capability: ABEX achieves higher mutation scores than the QD-based baseline and kills more hard-to-detect stubborn mutants. The analysis further highlights a complementary interplay between strategies: generation expands coverage, while mutation refines boundary quality. ABEX demonstrates that boundary value exploration can move from handcrafted, datatype-specific algorithms to adaptive, strategy-driven systems. This suggests a broader shift: LLM-guided search can serve as a general mechanism for automating complex testing tasks that were previously difficult to scale across input domains.

\section{Acknowledgments}
This work was supported by the Wallenberg AI, Autonomous Systems and Software Program (WASP), funded by the Knut and Alice Wallenberg Foundation. ChatGPT and Claude.ai was utilized to improve the phrasing of some parts of the text, originally written by the authors. The authors reviewed and edited the output and take full responsibility for the content of the article.

\bibliographystyle{elsarticle-num} 
\bibliography{references}
\end{document}